\documentclass[english]{article}
\usepackage[T1]{fontenc} \usepackage[latin9]{inputenc} 
\usepackage{amsthm}

\PassOptionsToPackage{normalem}{ulem} 
\usepackage{amsmath,amssymb,babel,cite,color,dsfont,empheq,framed,geometry,graphicx,mathrsfs,needspace,setspace,tabularx,tikz,ulem,url}
\newtheorem*{proposition}{Proposition (real wave velocities)}

\usepackage[capbesideposition=inside,
facing=yes,capbesidesep=quad]{floatrow}

\makeatletter
\newcommand{\vast}{\bBigg@{4}}
\newcommand{\Vast}{\bBigg@{12}}
\makeatother

\makeatletter

\let\@fnsymbol\@arabic

\makeatother

\newcommand{\Csym}{\mathbb{C}}

\newcommand{\mm}{\mu_{\mathrm{macro}}}
\newcommand{\lm}{\lambda_{\mathrm{macro}}}
\newcommand{\mh}{\mu_{\mathrm{micro}}}
\newcommand{\lh}{\lambda_{\mathrm{micro}}}
\newcommand{\me}{\mu_{e}}
\newcommand{\mc}{\mu_{c}}
\newcommand{\lle}{\lambda_{e}}
\newcommand{\ke}{\kappa_{e}}
\newcommand{\kh}{\kappa_{\mathrm{micro}}}
\newcommand{\km}{\kappa_{\mathrm{macro}}}
\newcommand{\mLc}{\me \,L_{c}^{2}}

\newcommand{\R}{\mathbb{R}}
\newcommand{\nablau}{\,\nabla u\,}
\newcommand{\p}{{P}}
\newcommand{\nablap}{\nabla \p}
\newcommand{\Curl}{\,\mathrm{Curl}}
\newcommand{\dev}{\, \mathrm{dev}}
\newcommand{\Div}{\mathrm{Div}}
\newcommand{\tr}{\, \mathrm{tr}\,}
\newcommand{\sym}{\, \mathrm{sym}\,}

\renewcommand{\skew}{\, \mathrm{skew}\,}

\newcommand{\so}{\mathfrak{so}}

\renewcommand{\skew}{\, \mathrm{skew}}

\newcommand{\devsym}{\dev\sym}

\newcommand{\id}{\,\mathds{1}}

\definecolor{Green}{rgb}{0,0.52,0}

\newcommand*\widefbox[1]{\fbox{\hspace{2em}#1\hspace{2em}}}

\title{\vspace{-1.0cm}Real wave propagation in the isotropic relaxed micromorphic model}

\author{
	Patrizio Neff\,\footnote{Patrizio Neff, corresponding author, patrizio.neff@uni-due.de, Head of Chair for 	Nonlinear Analysis and Modelling, Fakultät für Mathematik, Universität Duisburg-Essen,  Mathematik-Carrée, Thea-Leymann-Straße 9, 45127 Essen} \, and 
	Angela Madeo\footnote{Angela Madeo, angela.madeo@insa-lyon.fr, LGCIE, INSA-Lyon, Université de Lyon, 20 avenue	Albert Einstein, 69621, Villeurbanne cedex, France and IUF, Institut universitaire de France, 1 rue Descartes, 75231 Paris Cedex 05, France}  \, and 
	Gabriele Barbagallo\footnote{Gabriele Barbagallo, gabriele.barbagallo@insa-lyon.fr, LaMCoS-CNRS \& LGCIE, INSA-Lyon, Universitité de Lyon, 20 avenue Albert Einstein, 69621, Villeurbanne cedex, France} \\\, and 
	Marco Valerio 	d'Agostino\footnote{Marco Valerio d'Agostino, marco-valerio.dagostino@insa-lyon.fr, LGCIE, INSA-Lyon, Université de Lyon, 20 avenue Albert Einstein, 69621, Villeurbanne cedex, France} \ 
	and 	
	Rafael Abreu\footnote{Rafael Abreu, abreu@uni-muenster.de, Institut für Geophysik, Westfälische Wilhelms-Universität Münster, Corrensstraße 24, 48149, Münster,  Germany} \ 
	 and 
		Ionel-Dumitrel Ghiba\footnote{Ionel-Dumitrel Ghiba,  dumitrel.ghiba@uni-due.de, dumitrel.ghiba@uaic.ro,  Lehrstuhl für Nichtlineare Analysis und Modellierung, Fakultät für Mathematik, Universität Duisburg-Essen, Thea-Leymann Str.  9, 45127 Essen, Germany; Alexandru Ioan Cuza University of Ia{\c{s}}i, Department of Mathematics, Blvd.  Carol I, no.  11, 700506 Ia{\c{s}}i, Romania; and Octav Mayer Institute of Mathematics of the Romanian Academy, Ia{\c{s}}i Branch, 700505 Ia{\c{s}}i.
		} 
		}

\begin{document}

	\maketitle 
		\begin{abstract}
		For the recently introduced isotropic relaxed micromorphic generalized continuum model, we show that under the assumption of positive definite energy, planar harmonic waves have real velocity. We also obtain a necessary and sufficient condition for real wave velocity which is weaker than positive-definiteness of the energy. Connections to isotropic linear elasticity and micropolar elasticity are established. Notably, we show that strong ellipticity does not imply real wave velocity in micropolar elasticity, while it does in isotropic linear elasticity.
		\end{abstract}

		\addtocounter{footnote}{6} 
		
		\vspace{0.4cm}
		
		\hspace{-0.55cm}\textbf{Keywords}: ellipticity, positive-definiteness, real wave velocity, planar harmonic waves, rank-one convexity, acoustic tensor, generalized continuum, micropolar, Cosserat, micromorphic
		
		\vspace{0.4cm}
		
		\hspace{-0.55cm}\textbf{AMS 2010 subject classification}:  74A10 (stress), 74A30 (nonsimple materials), 74A60 (micromechanical theories),  74E15 (crystalline structure), 74M25 (micromechanics), 74Q15 (effective constitutive equations)
		
		
	\section{Introduction}
Investigations of real wave propagation and ellipticity are not new in principle. Indeed, it is textbook knowledge for linear elasticity that positive definiteness of the elastic energy implies real wave velocities (phase velocities) $v=\omega/k$  where $\omega\,[\mathrm{Hz}]$ is the angular frequency and $k[1/\mathrm{m}]\in\R$ is the wavenumber of planar propagating waves. In classical elasticity, having real wave velocities is equivalent to rank-one convexity (strong ellipticity or Legendre-Hadamard ellipticity). Moreover, ellipticity is equivalent to the positive definiteness of the acoustic tensor. For anisotropic  linear elasticity we mention \cite{chirita2007strong}, while for anisotropic nonlinear elasticity we refer  the reader to \cite{balzani2006polyconvex,merodio2006note,schroder2002application,schroder2008anisotropic}.

The same question of ellipticity and real wave velocities in generalized continuum mechanics has been discussed for micropolar models, e.g. in \cite{smith1967waves} and for elastic materials with voids in \cite{chirita2009strong}. For the isotropic micromorphic model results can be found with respect to positive definite energy and/or real wave velocity \cite{nowacky1985theory,smith1968inequalities}, Mindlin \cite{mindlin1963microstructure,mindlin1964micro} and Eringen's book \cite[pp. 277-280]{eringen1999microcontinuum}. These latter results present conditions which are neither easily verifiable nor are truly transparent. This is due to a certain lack of mathematical structure of the classical micromorphic model. Indeed, the implication that positive definiteness of the energy always implies real wave velocities is not directly established and demonstrated. In this paper we investigate the relaxed micromorphic model in terms of conditions for real wave velocities for planar waves and establish a necessary and sufficient conditions for this to happen.

This paper is organized as follows. We shortly recall the basics of the relaxed micromorphic model and discuss the wave propagation problem for propagating planar waves. Since we deal with an isotropic model, we can, without loss of generality, assume wave propagation in one specific direction only. The dispersion relations are then obtained and real wave-velocities under assumption of uniform-positiveness of the elastic energy are established.

We next present a set of necessary and sufficient conditions for real wave-velocities in the relaxed micromorphic model which is weaker than positivity of the energy, as the strong ellipticity condition is with respect to positive definiteness of the energy in the case of linear elasticity. Then, for didactic purposes, we repeat the analysis for isotropic linear elasticity in order to see relations of our necessary and sufficient condition to the strong ellipticity condition in linear elasticity. Similarly, we discuss micropolar elasticity and establish necessary and sufficient conditions for real wave propagation. We finally show that strong ellipticity in micropolar and micromorphic models is \textbf{not} sufficient for having real wave velocities, when dealing with plane waves. 
	
	\section{The relaxed micromorphic model}
		The relaxed micromorphic model has been recently introduced into continuum mechanics in \cite{neff2014unifying}. In subsequent works \cite{madeo2014band,madeo2015wave,madeo2016first,madeo2016reflection}, the model has shown its wider applicability compared to the classical Mindlin-Eringen micromorphic model in diverse areas \cite{abreu2016refined,eringen1999microcontinuum,greene2012microelastic,mindlin1963microstructure,mindlin1964micro}. 
		
		The dynamic relaxed micromorphic model counts only 8 constitutive parameters in the (simplified) isotropic case ($\me$, $\lle$, $\mh$, $\lh$, $\mc$, $L_c$, $\rho$, $\eta$). The simplification consists in assuming one scalar micro-inertia parameter $\eta$ and a uni-constant curvature expression. The characteristic length $L_c$ is intrinsically related to non-local effects due to the fact that it weights a suitable combination of first order space derivatives of the microdistorion tensor in the strain energy density \eqref{eq:Ener-2}. For a general presentation of the features of the relaxed micromorphic model in the anisotropic setting, we refer to \cite{barbagallo2016transparent}.
		\subsection{Elastic energy density\label{EN}}
	
	The relaxed micromorphic model couples the macroscopic displacement $ u\in\R^{3}$ and an affine substructure deformation attached at each macroscopic point encoded by the \textbf{micro-distortion} field $\p\in\R^{3\times3}$.	Our novel relaxed micromorphic model endows Mindlin-Eringen's representation of linear micromorphic models with the second order \textbf{dislocation density tensor}  $\alpha=-\Curl \p$ instead of the full gradient $\nablap$.\footnote{The dislocation tensor is defined as $\alpha_{ij}=-\left(\Curl \p \right)_{ij}=-\p_{ih,k}\epsilon_{jkh}$, where $\epsilon$ is the Levi-Civita tensor.} In the isotropic hyperelastic case the elastic energy reads
	\begin{align}
	W=&\ \me\,\lVert \sym\left(\nablau-\p\right)\rVert ^{2}+\frac{\lle}{2}\left(\mathrm{tr} \left(\nablau-\p\right)\right)^{2}+\mc\,\lVert \skew\left(\nablau-\p\right)\rVert ^{2}\label{eq:Ener-2}\\
	& 
	+\mh\,\lVert \sym \p\rVert ^{2}+\frac{\lh}{2}\,\left(\mathrm{tr} \p\right)^{2}
	+\frac{\mLc}{2} \,\lVert \Curl \p\rVert^2
	\nonumber  \\
			=&\ \underbrace{\me\,\lVert \devsym\left(\nablau-\p\right)\rVert ^{2}+\frac{2\,\me+3\,\lle}{3}\left(\mathrm{tr} \left(\nablau-\p\right)\right)^{2}}_{\mathrm{{\textstyle isotropic\ elastic-energy}}}	+\hspace{-0.1cm}\underbrace{\mc\,\lVert \skew\left(\nablau-\p\right)\rVert ^{2}}_{\mathrm{\textstyle rotational\   elastic\ coupling}		}\hspace{-0.1cm} \nonumber\\\nonumber
		& 
		+\underbrace{\mh\,\lVert \devsym \p\rVert ^{2}+\frac{2\,\mh+3\,\lh}{3}\,\left(\mathrm{tr}\, \p\right)^{2}}_{\mathrm{{\textstyle micro-self-energy}}}
		+\hspace{-0.95cm}\underbrace{\frac{\mLc}{2} \,\lVert \Curl\, \p\rVert^2}_{\mathrm{\textstyle simplified\ isotropic\ curvature}}\,,
	\end{align}
	where the parameters and the elastic stress are analogous to the standard Mindlin-Eringen micromorphic model. The model is well-posed in the statical and dynamical case even for zero Cosserat couple modulus $\mc=0$, see \cite{neff2015relaxed,ghiba2014relaxed}. In that case, it is non-redundant in the sense of \cite{romano2016micromorphic}. Well-posedness results for the statical and dynamical cases have been provided in \cite{neff2014unifying} making decisive use of recently established new coercive inequalities, generalizing Korn's inequality to incompatible tensor fields \cite{neff2015poincare,neff2012maxwell,neff2011canonical,bauer2014new,bauer2016dev}.
	
	Strict positive definiteness of the potential energy is equivalent to the following simple relations for the introduced parameters \cite{neff2014unifying}:
		\begin{equation}
		\me>0,\qquad\mc>0,\qquad 2\,\me+3\,\lle>0,\qquad\mh>0,\qquad2\,\mh+3\,\lh>0,\qquad L_c>0.\label{DefPos}
		\end{equation}
As for the kinetic energy, we consider that it takes the following (simplified) form 
	\begin{gather}
	J=\frac{\rho}{2}\left\Vert u_{,t}\right\Vert ^{2}+\hspace{-1.2cm}\underbrace{\frac{\eta}{2}\left\Vert \p_{,t}\right\Vert ^{2},}_{\mathrm{\textstyle simplified\ micro-inertia}}\label{eq:Kinetic_energy2}
	\end{gather}
		where $\rho>0$ is the value of the averaged macroscopic mass density of the considered material, while $\eta>0$ is its micro-inertia density.

For very large sample sizes, a scaling argument shows easily
that the relative characteristic length scale $L_{c}$ of
the micromorphic model must vanish. Therefore, we have a way of comparing a classical first gradient formulation with the relaxed micromorphic model 
and to offer an a priori relation between the microscopic parameters $\lle,\lh,\me,\mh$
on the one side and the resulting macroscopic parameters $\lm,\mm$ on the other side \cite{barbagallo2016transparent,neff2004material,neff2007geometrically}. We have
\begin{align}
\left(2\mm+3\lm\right)= &\, \frac{\left(2\me+3\lle\right)\left(2\mh+3\lh\right)}{\left(2\me+3\lle\right)+\left(2\mh+3\lh\right)},\label{eq:IsotropicRel-2}\qquad \qquad\qquad \mm=\,\frac{\me\,\mh}{\me+\mh},
\end{align}
where $\mm,\lm$ are the moduli obtained for $L_c\rightarrow0$. 

		For future use we define the elastic bulk modulus $\ke$, the microscopic bulk modulus $\kh$ and the macroscopic bulk modulus $\km$, respectively:
\begin{align}
\ke=\frac{2\,\me+3\,\lle}{3},\qquad\qquad\kh=\frac{2\,\mh+3\,\lh}{3},\qquad\qquad\km=\frac{2\,\mm+3\,\lm}{3}.
\end{align}		
In terms of these moduli, strict positive-definiteness of the energy is equivalent to:
		\begin{equation}
		\me>0,\qquad\mc>0,\qquad \ke>0,\qquad\mh>0,\qquad \kh>0,\qquad L_c>0.\label{Pos}
		\end{equation}
		If strict positive-definiteness \eqref{Pos} holds we can write the macroscopic consistency conditions as:
\begin{align}
\km= &\, \frac{\ke \,\kh}{\ke+\kh},&\mm&=\,\frac{\me	\,\mh}{\me+\mh},
\end{align}	
and, again under condition \eqref{Pos}
		\begin{align}
		 \ke=&\, \frac{\kh \,\km}{\kh-\km},
		&\kh=&\, \frac{\ke \,\km}{\ke-\km},
		&\me&=\,\frac{\mh	\,\mm}{\mh-\mm},
		&\mh&=\,\frac{\me	\,\mm}{\me-\mm}.
		\end{align}	
Here, strict positivity \eqref{Pos} implies that:
\begin{align}
\ke+\kh&>0,&\me+\mh&>0,&\ke&>\km,&\kh&>\km,&\me>&\mm,\\\nonumber \mh&>\mm.
\end{align} 
Since it is useful in what follows we explicitly remark that:
\begin{align}
2\,\me+\lle&=\frac{4}{3}\,\me+\frac{2\,\me+3\,\lle}{3}=\frac{4}{3} \,\me+\ke=\frac{4 \,\me+3\,\ke}{3},\qquad  2\,\mh+\lh=\frac{4\,\mh+3\,\kh}{3}.
\end{align}
With these relationship, it is easy to show how $	\me>0$ and $\ke>0$ imply $2\,\me+\lle>0$. Moreover,  as shown in the appendix (equations \eqref{dis1} and \eqref{dis2}), we note here that if only $\me+\mh>0$ and $\ke+\kh>0$, then the macroscopic parameters are less or equal than respective microscopic parameters, namely:
\begin{align}
\ke\geq\km,\qquad\qquad\kh\geq\km\qquad\qquad \me\geq\mm,\qquad\qquad \mh\geq\mm,
\end{align} 
and moreover the following inequalities are satisfied:
\begin{align}
2\,\me+\lle&\geq 2\,\mm+\lm,&
2\,\mh+\lh&\geq 2\,\mm+\lm,&
\frac{4\,\mm+3\,\ke}{3}&\geq2\,\mm+\lm.
\end{align} 
Note that the Cosserat couple modulus $\mc$ \cite{neff2006cosserat} does not appear in the introduced scale between micro and macro.

\subsection{Dynamic formulation}
\begin{figure}[!h]
	\begin{centering}
		\begin{tabular}{ccccc}
			\includegraphics[width=5cm]{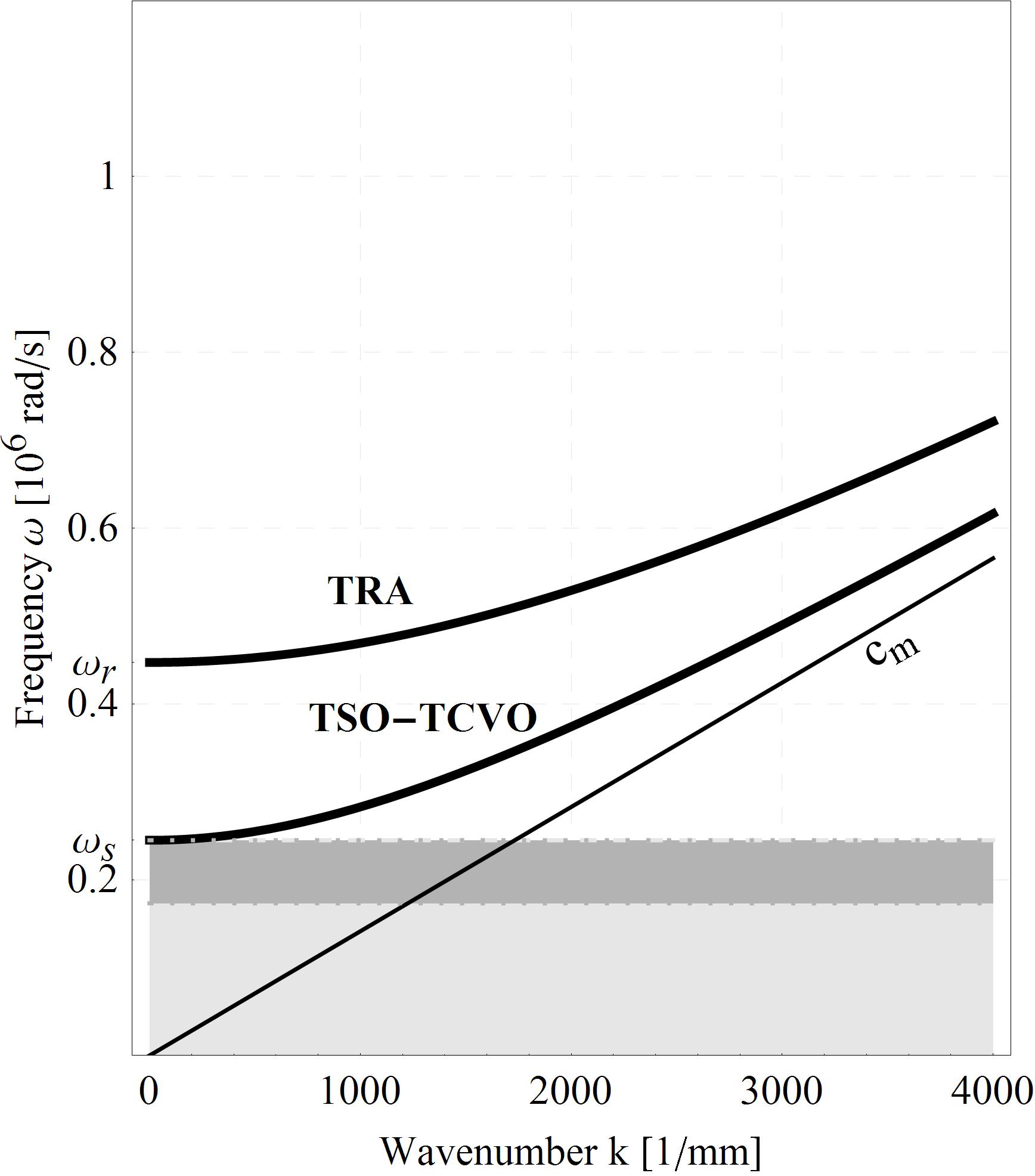}  &
			\includegraphics[width=5cm]{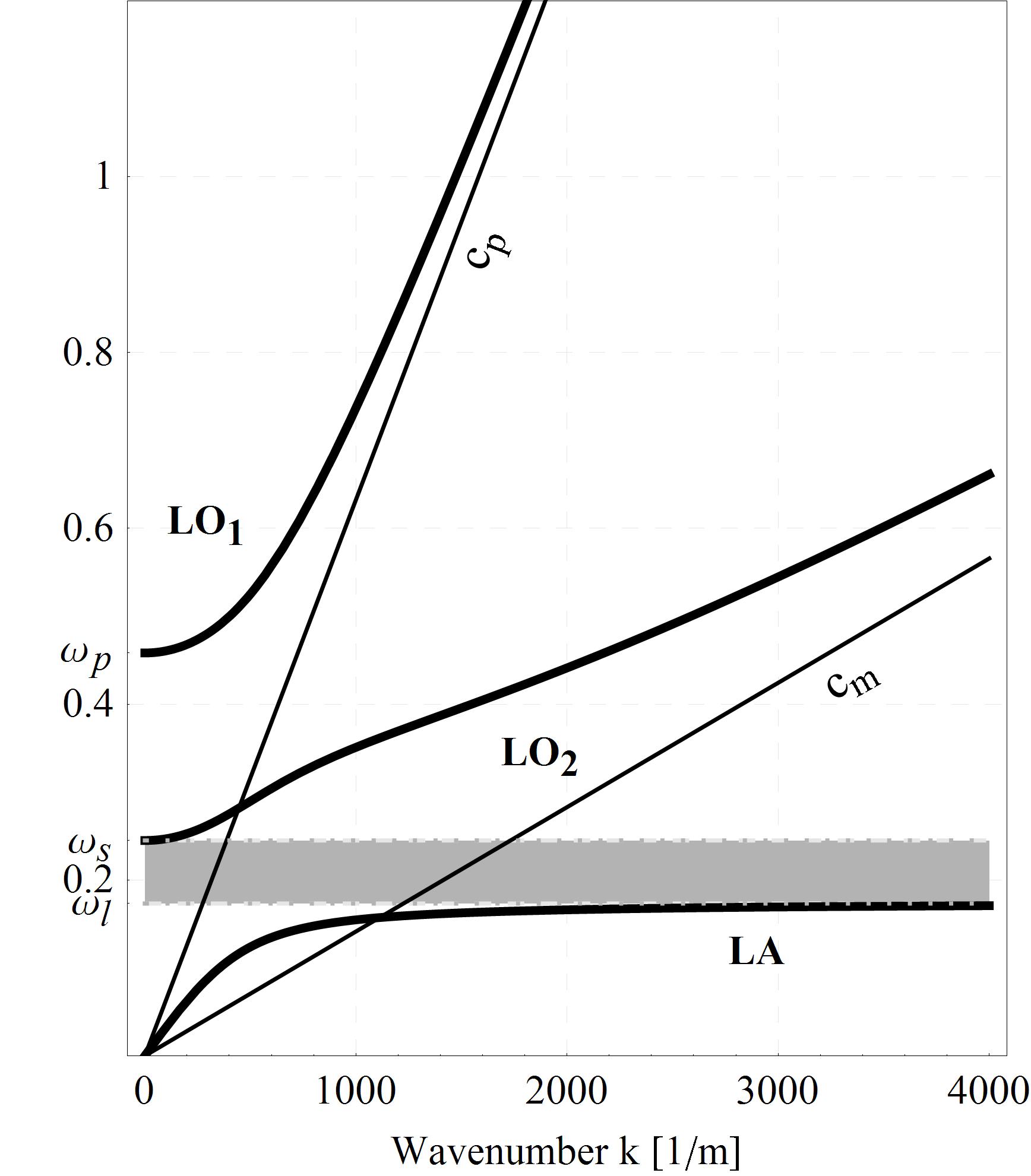} &
			\includegraphics[width=5cm]{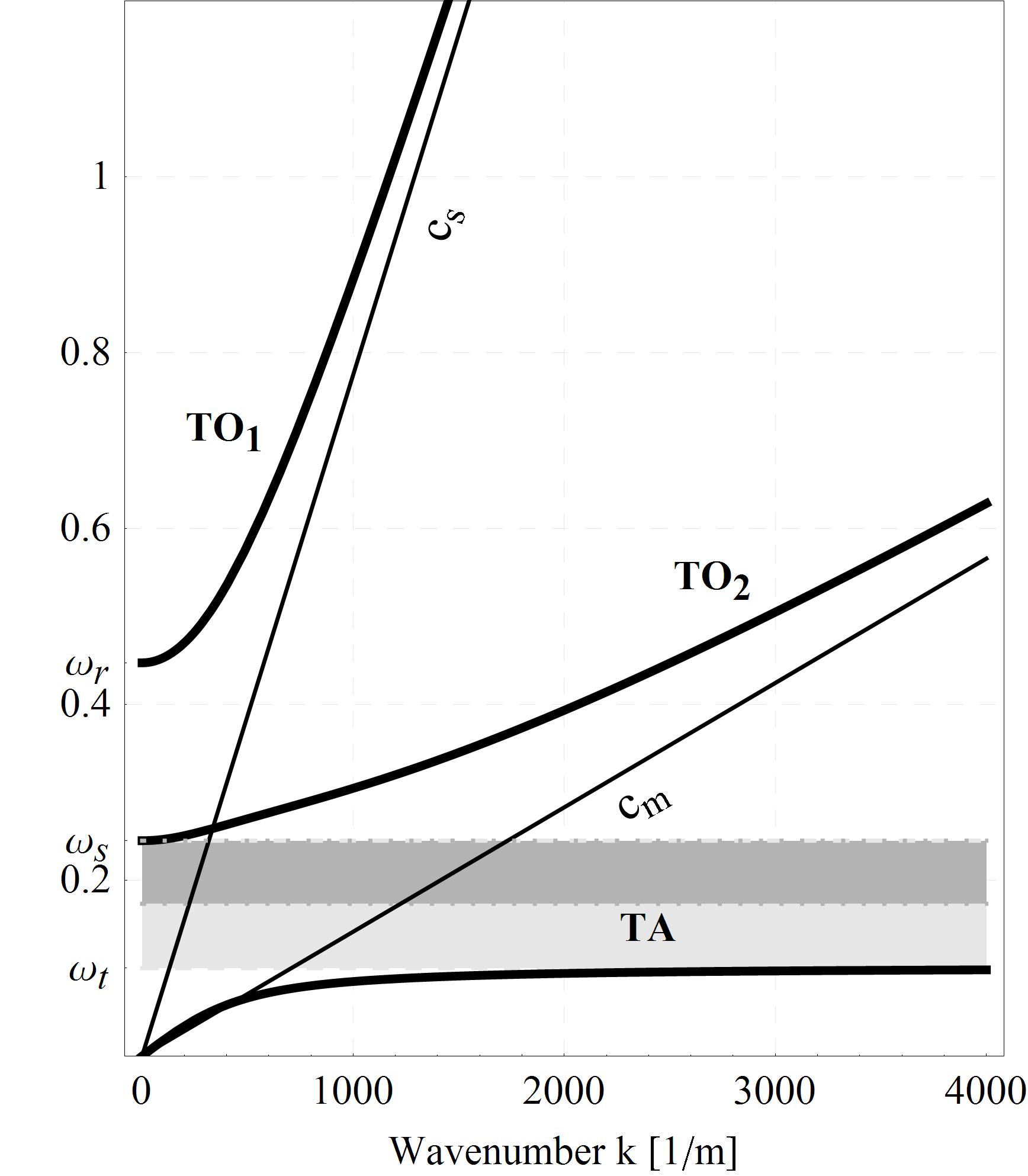} \\
			\quad(a) $\mathrm{det}\,\mathbf{\overline{A}}_{4}(\omega,k)=0$&\quad(b) $\mathrm{det}\,\mathbf{\overline{A}}_{1}(\omega,k)=0$&\quad(c) $\mathrm{det}\,\mathbf{\overline{A}}_{2}(\omega,k)=0$
		\end{tabular}
		\par\end{centering}
	
	\caption{\label{CurlNon}Dispersion relations $\omega=\omega(k)$ \cite{madeo2016first} for the
		\textbf{relaxed micromorphic model}  with non-vanishing Cosserat couple modulus $\mc>0$.  Uncoupled waves (a), longitudinal waves (b) and
		transverse waves (c). TRO: transverse rotational optic, TSO: transverse shear optic, TCVO:
		transverse constant-volume optic, LA: longitudinal acoustic, LO$_{1}$-LO$_{2}$:
		$1^{st}$ and $2^{nd}$ longitudinal optic, TA: transverse acoustic, TO$_{1}$-TO$_{2}$: $1^{st}$ and $2^{nd}$ transverse optic.
	}	
\end{figure}

The dynamical formulation is obtained  defining a joint Hamiltonian and assuming stationary action.  The dynamical equilibrium equations are:
\begin{align}
\rho\,  u_{,tt}=&\,\Div\left[2\,\me\, \sym\left(\nablau-\p\right)+2\,\mc \,\skew\left(\nablau-\p\right)+\lle\tr\left(\nablau-\p\right)\mathds{1}\right] , \nonumber \\
\eta  \p_{,tt}=&\,-\mLc  \Curl  \Curl \,\p+2\, \me \,\sym\left(\nablau-\p\right) +2 \,\mc\, \skew\left(\nablau-\p\right)\label{eq:Dyn}\\&\hspace{3.075cm} +\lle\tr\left(\nablau-\p\right)\mathds{1}-\left[2\,\mh \sym \p +\lh \tr(\p)\mathds{1}\right].\nonumber 
\end{align}
Sufficiently far from a source, dynamic wave solutions may be treated as planar waves. Therefore, we now want to study harmonic solutions traveling in an infinite domain for the differential system
\eqref{eq:Dyn}. To do so, we define:
\begin{align}
\p^{S}& := \frac{1}{3}\tr\left( \p\right),&\p_{\left[ij\right]}&:=\left(\skew \p \right)_{ij}= \frac{1}{2}\left(\p_{ij}-\p_{ji}\right), \label{Decom}\\\nonumber
\p^{D}&:=\p_{11}- \p^S, &\p_{\left(ij\right)}
&:= \left(\sym\p \right)_{ij}=\frac{1}{2}\left(\p_{ij}+\p_{ji}\right),\\\nonumber
P^{V}&:=P_{22}-P_{33}
\end{align}
and we introduce the unknown vectors
\begin{align} \mathbf{v}_{1}=\left(u_{1},P^{D},P^{S}\right)\qquad\qquad\mathbf{v}_{\tau}=\left(u_{\tau},P_{(1\tau)},P_{[1\tau]}\right),\qquad\qquad\tau=2,3,\qquad\qquad \mathbf{v}_{4}=\left(P_{(23)},P_{[23]},P^{V}\right).
\end{align}
 We suppose that the space dependence of all introduced kinematic fields are limited to a direction defined by a unit vector $\widetilde{\xi}\in\R^3$, which is the direction of propagation of the wave. Hence, we look for solutions of \eqref{eq:Dyn} in the form:
\begin{equation}
\underbrace{\mathbf{v}_{1}= \boldsymbol{\beta}	\,e^{i(k\langle \widetilde{\xi},\, x\rangle_{\R^3}-\omega t)}}_{\text{longitudinal}} ,\qquad\underbrace{\mathbf{v}_{\tau}= \boldsymbol{\gamma}	\,^{\tau}e^{i(k\langle \widetilde{\xi},\, x\rangle_{\R^3}-\omega t)}}_{\text{transversal}},\qquad \tau=2,3,\qquad\underbrace{\mathbf{v}_{4}= \boldsymbol{\gamma}	\,^{4}e^{i(k\langle \widetilde{\xi},\, x\rangle_{\R^3}-\omega t)}}_{\text{uncoupled}}.\label{WaveForm1}
\end{equation}
Since our formulation is isotropic, we can, without loss of generality, specify the direction $\widetilde{\xi}=e_1$.   Then  $X=\langle e_1,x\rangle=x_1$, and we obtain that the space dependence of all introduced kinematic fields are limited to the component $X$ which is the direction of propagation of the wave\footnote{In an isotropic model it is clear that there is no direction dependence. More specifically, let us consider an arbitrary direction  $\widetilde{\xi}\in\R^3$. Now we consider an orthogonal spatial coordinate change $Q\,e_1=\widetilde{\xi}$ with $Q\in\mathrm{SO}(3)$. In the rotated variables, the ensuing system of pde's \eqref{eq:Dyn} is form-invariant, see \cite{munch2016rotational}.}. This means that we look for solutions in the form:
\begin{equation}
\underbrace{\mathbf{v}_{1}= \boldsymbol{\beta}	\,e^{i(kX-\omega t)}}_{\text{longitudinal}} ,\qquad\qquad\underbrace{\mathbf{v}_{\tau}= \boldsymbol{\gamma}	\,^{\tau}e^{i(kX-\omega t)}}_{\text{transversal}},\qquad\qquad \tau=2,3,\qquad\qquad\underbrace{\mathbf{v}_{4}= \boldsymbol{\gamma}	\,^{4}e^{i(kX-\omega t)}}_{\text{uncoupled}} ,\label{WaveForm2}
\end{equation}
where $\boldsymbol{\beta}=(\beta_{1},\beta_{2},\beta_{3})^{T}\in\mathbb{C}^{3}$,
$\boldsymbol{\gamma}^{\tau}=(\gamma_{1}^{\tau},\gamma_{2}^{\tau},\gamma_{3}^{\tau})^{T}\in\mathbb{C}^{3}$ and
$\boldsymbol{\gamma}^{4}=(\gamma_{1}^{4},\gamma_{2}^{4},\gamma_{3}^{4})^{T}\in\mathbb{C}^{3}$
are the unknown amplitudes of the considered waves\footnote{Here, we understand that having found the (in general, complex) solutions of \eqref{WaveForm2} only the real or imaginary parts separately constitute actual wave solutions which can be observed in reality.}, $k$ is the wavenumber and $\omega$ is the wave-frequency. Replacing these expressions in equations \eqref{eq:Dyn}, it is possible to express the system (see \cite{madeo2014band,madeo2015wave}) as:
\begin{equation}
\mathbf{A}_{1}\cdot\boldsymbol{\beta}=0,\qquad\qquad\mathbf{A}_{\tau}\cdot\boldsymbol{\gamma}^{\tau}=0,\qquad\qquad\tau=2,3,\qquad\qquad\mathbf{A}_{4}\cdot\boldsymbol{\gamma}^{4}=0,\label{AlgSys}
\end{equation}
with
\begin{align}
\mathbf{A}_{1}(\omega,k)\,&=\left(\begin{array}{ccc}
-\omega^{2}+c_{p}^{2}\, k^{2} & \, i\: k\:2\me/\rho\  & i\: k\:\left(2\,\me+3\,\lle\right)/\rho\\
\\
-i\: k\,\frac{4}{3}\,\me/\eta & -\omega^{2}+\frac{1}{3}k^{2}c_{m}^{2}+\omega_{s}^{2} & -\frac{2}{3}\, k^{2}c_{m}^{2}\\
\\
-\frac{1}{3}\, i\, k\:\left(2\,\me+3\,\lle\right)/\eta & -\frac{1}{3}\, k^{2}\, c_{m}^{2} & -\omega^{2}+\frac{2}{3}\, k^{2}\, c_{m}^{2}+\omega_{p}^{2}
\end{array}\right),
\\\nonumber\\
\mathbf{A}_{2}(\omega,k)=\mathbf{A}_{3}(\omega,k)\,&=\left(\begin{array}{ccc}
-\omega^{2}+k^{2}c_{s}^{2}\  & \, i\, k\,2\me/\rho\  & -i\, k\,\frac{\eta}{\rho}\omega_{r}^{2},\\
\\
-\, i\, k\,\me/\eta, & -\omega^{2}+\frac{c_{m}^{2}}{2}k^{2}+\omega_{s}^{2} & \frac{c_{m}^{2}}{2} k^{2}\\
\\
\frac{i}{2}\,\omega_{r}^{2}\, k &\frac{c_{m}^{2}}{2} k^{2} & -\omega^{2}+\frac{c_{m}^{2}}{2} k^{2}+\omega_{r}^{2}
\end{array}\right),
\\\nonumber\\
\mathbf{A}_{4}(\omega,k)\,&=\left(\begin{array}{ccc}
-\omega^{2}+c_{m}^{2}\, k^{2}+\omega_s^2 & 0 & 0\\
\\
0& -\omega^{2}+c_{m}^{2}\, k^{2}+\omega_r^2 & 0\\
\\
0 & 0& -\omega^{2}+c_{m}^{2}\, k^{2}+\omega_s^2 
\end{array}\right).
\end{align}
Here, we have defined:
\begin{empheq}[box=\widefbox]{align}
c_{m}&=\sqrt{\frac{\mLc}{\eta}},\qquad& c_{s}&=\sqrt{\frac{\me+\mc}{\rho}},\qquad &c_{p}&=\sqrt{\frac{2\,\me+\lle}{\rho}},\nonumber \\
\nonumber \\
\omega_{s}&=\sqrt{\frac{2\left(\me+\mh\right)}{\eta}},\qquad
&\omega_{p}&=\sqrt{\frac{\left(2\,\me+3\,\lle\right)+\left(2\,\mh+3\,\lh\right)}{\eta}},\qquad&\omega_{r}&=\sqrt{\frac{2\,\mc}{\eta}},\nonumber\\
\nonumber \\
\omega_{l}&=\sqrt{\frac{2\,\mh+\lh}{\eta}},\qquad&\omega_{t}&=\sqrt{\frac{\mh}{\eta}}.\nonumber 
\end{empheq}
Let us next define the diagonal matrix:
\begin{align}
\mathrm{diag}_1=\left(\begin{array}{ccc}\sqrt{\rho}&0&0\\0&i \frac{\sqrt{6\eta}}{2} &0\\0&0& i \sqrt{3\eta}\end{array}\right).
\end{align}
Considering $\gamma=\mathrm{diag}_1\cdot\beta$ and the matrix $\overline{\mathbf{A}}_{1}(\omega,k)=\mathrm{diag}_1\cdot\mathbf{A}_{1}(\omega,k)\cdot\mathrm{diag}_1^{-1}$, it is possible to formulate the problem \eqref{AlgSys} equivalently as\footnote{It is possible to face the problem in two more equivalent ways. The first one is to consider from the start that the amplitudes of the micro-distortion field are multiplied by the imaginary unit i, i.e.  $\boldsymbol{\beta}=(\beta_{1},i\, \beta_{2},i\, \beta_{3})^{T}\in\mathbb{C}^{3}$, as done in \cite[p. 24, eq. 8.6]{mindlin1963microstructure}. Doing so, we obtaining a real matrix that can be symmetrized with $\mathrm{diag}_1=\left(\begin{array}{ccc}\sqrt{\rho}&0&0\\0& \frac{\sqrt{6\eta}}{2} &0\\0&0& \sqrt{3\eta}\end{array}\right)$. On the other hand, it is also possible to consider from the beginning  $\boldsymbol{\beta}=(\sqrt{\rho} \beta_{1},i\frac{\sqrt{6\eta}}{2}\, \beta_{2},i \sqrt{3\eta}\, \beta_{3})^{T}\in\mathbb{C}^{3}$ obtaining directly a real symmetric matrix. }: 

\begin{align}
\overline{\mathbf{A}}_{1}\cdot\gamma=\left(\begin{array}{ccc}
-\omega^{2}+c_{p}^{2}\, k^{2} & \frac{2\sqrt{6}}{3}\: k\,\me/\sqrt{\rho\eta} &\frac{\sqrt{3}}{3}\, k\:\left(2\,\me+3\,\lle\right)/\sqrt{\rho\eta} \\
\\
\frac{2\sqrt{6}}{3}\: k\,\me/\sqrt{\rho\eta} & -\omega^{2}+\frac{1}{3}k^{2}c_{m}^{2}+\omega_{s}^{2} & -\frac{\sqrt{2}}{3}\, k^{2}c_{m}^{2}\\
\\
\frac{\sqrt{3}}{3}\, k\:\left(2\,\me+3\,\lle\right)/\sqrt{\rho\eta} & -\frac{\sqrt{2}}{3}\, k^{2}\, c_{m}^{2} & -\omega^{2}+\frac{2}{3}\, k^{2}\, c_{m}^{2}+\omega_{p}^{2}
\end{array}\right)\left(\begin{array}{c}\gamma_1\\\gamma_2\\\gamma_3\end{array}\right)=0.
\end{align}
Analogously considering 
\begin{align}
\mathrm{diag}_2=\left(\begin{array}{ccc}\sqrt{\rho}&0&0\\0&i \sqrt{2\eta} &0\\0&0& i \sqrt{2\eta}\end{array}\right),
\end{align}
it is possible to obtain  $\overline{\mathbf{A}}_{2}(\omega,k)=\overline{\mathbf{A}}_{3}(\omega,k)=\mathrm{diag}_2\cdot\mathbf{A}_{2}(\omega,k)\cdot\mathrm{diag}_2^{-1}$
\begin{align}
\mathbf{\overline{A}}_{2}(\omega,k)=\mathbf{\overline{A}}_{3}(\omega,k)\,&=\left(\begin{array}{ccc}
-\omega^{2}+k^{2}c_{s}^{2}\  & \, k\,\sqrt{2} \me/\sqrt{\rho\eta}\  & -k\,\sqrt{2}\,\mc/\sqrt{\rho\eta},\\
\\
k\,\sqrt{2} \me/\sqrt{\rho\eta}, & -\omega^{2}+\frac{c_{m}^{2}}{2}k^{2}+\omega_{s}^{2} & \frac{c_{m}^{2}}{2} k^{2}\\
\\
-k\,\sqrt{2}\,\mc/\sqrt{\rho\eta} &\frac{c_{m}^{2}}{2} k^{2} & -\omega^{2}+\frac{c_{m}^{2}}{2} k^{2}+\omega_{r}^{2}
\end{array}\right).
\end{align}
In order to have non-trivial solutions of the algebraic systems (\ref{AlgSys}),
one must impose that 
\begin{equation}
\mathrm{det}\,\mathbf{\overline{A}}_{1}(\omega,k)=0,\qquad\qquad\mathrm{det}\,\mathbf{\overline{A}}_{2}(\omega,k)=\mathrm{det}\,\mathbf{\overline{A}}_{3}(\omega,k)=0,\qquad\qquad\mathrm{det}\,\mathbf{A}_{4}(\omega,k)=0,\label{Dispersion}
\end{equation}
the solution of which allow us to determine the so-called dispersion relations $\omega=\omega\left(k\right)$ for the longitudinal and transverse waves in the relaxed micromorphic continuum, see Figure \ref{CurlNon}\footnote{The formal limit $\eta\rightarrow+\infty$ shows no dispersion at all giving two pseudo-acoustic linear curves, longitudinal and transverse with slopes $c_p=\sqrt{(2\me+\lle)/\rho}$ and $c_s=\sqrt{(\me+\mc)/\rho}$, respectively.}.
\newline 
For solutions $\omega=\omega(k)$ of \eqref{Dispersion} we define the
\begin{align}
\textbf{phase velocity: } v=\frac{\omega}{k}, \qquad\qquad \textbf{group velocity: }\ \frac{d\omega(k)}{dk}.
\end{align}
Real wave numbers $k\in\R$ correspond to propagating waves, while complex values of $k$ are associated with waves whose amplitude either grows or decays along the coordinate $X$. In linear elasticity, phase velocity and group velocity coincide since there is no dispersion and both are real, see section \ref{ClassLin}.\smallskip \newline
\boxed{
\begin{minipage}{\columnwidth}
Since in this paper we are only interested in real $k$, the wave velocity (phase velocity) is real if and only if $\omega$ is real.
\end{minipage}
}
\smallskip

Since $\omega^2$ appears on the diagonal only, the problem \eqref{Dispersion} can be analogously expressed as an eigenvalue-problem:
\begin{align}
\mathrm{det}\,\left(\mathbf{B}_{1}(k)-\omega^2\,\mathds{1}\right)&=0,&\mathrm{det}\,\left(\mathbf{B}_{2}(k)-\omega^2\,\mathds{1}\right)&=0,\label{Eig}\\\nonumber
\mathrm{det}\,\left(\mathbf{B}_{3}(k)-\omega^2\,\mathds{1}\right)&=0,&\mathrm{det}\,\left(\mathbf{B}_{4}(k)-\omega^2\,\mathds{1}\right)&=0,
\end{align}
where
\begin{align}
\mathbf{B}_{1}(k)\,&=\left(\begin{array}{ccc}
c_{p}^{2}\, k^{2} & \frac{2\sqrt{6}}{3}\: k\,\me/\sqrt{\rho\eta} &\frac{\sqrt{3}}{3}\,  k\:\left(2\,\me+3\,\lle\right)/\sqrt{\rho\eta} \\
\\
\frac{2\sqrt{6}}{3}\: k\,\me/\sqrt{\rho\eta} & \frac{1}{3}k^{2}c_{m}^{2}+\omega_{s}^{2} & -\frac{\sqrt{2}}{3}\, k^{2}c_{m}^{2}\\
\\
\frac{\sqrt{3}}{3}\, k\:\left(2\,\me+3\,\lle\right)/\sqrt{\rho\eta} & -\frac{\sqrt{2}}{3}\, k^{2}\, c_{m}^{2} & +\frac{2}{3}\, k^{2}\, c_{m}^{2}+\omega_{p}^{2}
\end{array}\right),\\\nonumber\\
\mathbf{B}_{2}(k)=\mathbf{B}_{3}(k)\,&=\left(\begin{array}{ccc}
k^{2}c_{s}^{2}\  &  k\,\sqrt{2} \me/\sqrt{\rho\eta}\  &-k\,\sqrt{2}\,\mc/\sqrt{\rho\eta},\\
\\
k\,\sqrt{2} \me/\sqrt{\rho\eta}, & \frac{c_{m}^{2}}{2}k^{2}+\omega_{s}^{2} & \frac{c_{m}^{2}}{2} k^{2}\\
\\
-k\,\sqrt{2}\,\mc/\sqrt{\rho\eta} &\frac{c_{m}^{2}}{2} k^{2} & \frac{c_{m}^{2}}{2} k^{2}+\omega_{r}^{2}
\end{array}\right),
\\\nonumber\\
\mathbf{B}_{4}(k)\,&=\left(\begin{array}{ccc}
c_{m}^{2}\, k^{2}+\omega_s^2 & 0 & 0\\
\\
0& c_{m}^{2}\, k^{2}+\omega_r^2 & 0\\
\\
0 & 0& c_{m}^{2}\, k^{2}+\omega_s^2 
\end{array}\right).
\end{align}
Note that $\mathbf{B}_{1}(k)$, $\mathbf{B}_{2}(k)$, $\mathbf{B}_{3}(k)$ and $\mathbf{B}_{4}(k)$ are real symmetric matrices and therefore the resulting eigenvalues $\omega^2$ are real. Obtaining real wave velocities is tantamount to having $\omega^2\ge0$ for all solutions of \eqref{Eig}.

\subsection{Necessary and sufficient conditions for real wave propagation}
We will show next that all the eigenvalues $\omega^2$ of $\mathbf{B}_{1}(k)$, $\mathbf{B}_{2}(k)$ and $\mathbf{B}_{3}(k)$ are real and positive for every $k\neq0$ and non-negative for $k=0$ provided certain conditions on the material coefficients are satisfied. Sylvester's criterion states that a Hermitian matrix M is positive-definite if and only if the leading principal minors are positive \cite{gilbert1991positive}. For the matrix $\mathbf{B}_{1}$ the three principal minors are:
\begin{align}
\left(\mathbf{B}_{1}\right)_{11}=&\ \frac{2\me+\lle}{\rho}, \\
(\mathrm{Cof}\,\left(\mathbf{B}_{1}\right))_{33}=&\frac{k^2}{3\eta \rho} \left[ 6(2\,\me+\lle) \mh +  6\, \me \, \ke+(2\,\me+\lle) \mLc k^2  \right]
\\\nonumber
=&\ \frac{k^2}{3\eta \rho} \left[2 \left(  4 \mm+3\ke \right) (\me+\mh) + (2\,\me+\lle) \mLc k^2 \right],\nonumber
\\\det\left(\mathbf{B}_{1}\right)=&\ \frac{k^2}{\eta^2 \rho} \bigg[6\,\ke \,\kh\,  (\me + \mh)+  8\, \me \mh (\ke+\kh) +(2\,\me+\lle) (2\,\mh+\lh) \mLc\, k^2 \bigg]
\\=&\ \frac{k^2}{\eta^2 \rho} \Big[6 \,(\ke+\kh)\big(\me + \mh\big)\big(2\,\mm+\lm\big)+\big(2\,\me+\lle\big) \big(2\,\mh+\lh\big)\mLc\, k^2\Big]. \nonumber
\end{align}
The three principal minors of $\mathbf{B}_{1}$ are clearly positive for $k\neq0$ if\footnote{We note here that $4\,\mm+3\,\ke>0\iff2\,\me+\lle>\frac{4}{3}(\me-\mm) \iff 2\,\mm+\lm>\km-\ke$. Furthermore, if $\me+\mh>0$ and $\ke+\kh>0$, we have $3\,(2\,\me+\lle)\geq4\,\mm+3\,\ke\geq3\,(2\,\mm+\lm)$, see Appendix.}:
\begin{align}
\me&>0,&\mh&>0, &\ke+\kh&>0,  &2\,\mm+\lm&>0, \label{suff} \\\nonumber 4\,\mm+3\,\ke&>0,& 2\,\me+\lle&>0, &2\,\mh+\lh&>0.
\end{align}
Similarly, for the matrix $\mathbf{B}_{2}$ the three principal minors are:
\begin{align}
\left(\mathbf{B}_{2}\right)_{11}=&\ \frac{\me+\mc}{\rho}, \\ 
(\mathrm{Cof}\,\left(\mathbf{B}_{2}\right))_{33}=&\ \frac{k^2}{2 \eta \rho} \Big[ 4\, (\me\, \mc+\mh(\me+\mc)+ (\me + \mc) \, \mLc k^2  \Big].
\\\det\,\left(\mathbf{B}_{2}\right)=&\ \frac{k^2}{\eta^2 \rho} \Big[ 4\, \mh\, \mc\, \me + (\me + \mc) \mh\, \mLc k^2  \Big].
\end{align}
For the matrix $\mathbf{B}_{2}(k)=\mathbf{B}_{3}(k)$, considering positive $\eta,\,\rho$ and separating terms in the brackets by looking at large and small values of $k$, we can state \textbf{necessary} and \textbf{sufficient} conditions for strict positive-definiteness of $\mathbf{B}_{2}(k)$ at arbitrary $k\neq 0$:
\begin{align}
\me>0,\qquad\qquad \mh>0,\qquad\qquad \mc\geq0. \label{NecTran}
\end{align}
Since $\mathbf{B}_{4}(k)$ is diagonal, it easy to show that positive definiteness is tantamount to the set of \textbf{necessary} and \textbf{sufficient} conditions for $k\neq0$:
 \begin{align}
 \me>0,\qquad\qquad \me+\mh>0,\qquad\qquad \mc\geq0. \label{NecUnc}
 \end{align}
 On the other hand, considering the case $k=0$, we obtain that the matrices reduce to:
 \begin{align}
 \mathbf{B}_{1}(0)\,&=\left(\begin{array}{ccc}
 0& 0 &0 
 \\0& \omega_{s}^{2} & 0
 \\0& 0 & \omega_{p}^{2}
 \end{array}\right),&
 \mathbf{B}_{2}(0)=\mathbf{B}_{3}(0)\,&=\left(\begin{array}{ccc}
 0&  0& 0
 \\
 0&\omega_{s}^{2} & 0
 \\
0 &0&\omega_{r}^{2}
 \end{array}\right),&
 \mathbf{B}_{4}(0)\,&=\left(\begin{array}{ccc}
 \omega_s^2 & 0 & 0
 \\
 0& \omega_r^2 & 0\\
 0 & 0& \omega_s^2 
 \end{array}\right).
 \end{align}
 Since the matrices are diagonal for $k=0$, it easy to show that positive semi-definiteness is tantamount to the set of \textbf{necessary} and \textbf{sufficient} conditions :
 \begin{align}
 \me\geq0,\qquad\qquad \me+\mh\geq0,\qquad\qquad \mc\geq0,\qquad\qquad \ke+\kh\geq0. \label{NecSemi}
 \end{align}
Hence, we can state a simple \textbf{sufficient} condition for real wave velocities for all real k:
\begin{align}
\me&>0,&\mh&>0, &\ke+\kh&>0,  &2\,\mm+\lm&>0, \\\nonumber 4\,\mm+3\,\ke&>0,& 2\,\me+\lle&>0, &2\,\mh+\lh&>0.
\end{align}
In order to see a set of global necessary conditions for positivity at arbitrary $k\neq0$ we consider first large and small values of $k\neq0$ separately. For $k\rightarrow+\infty$ we must have:
\begin{align}
2\,\me+\lle&>0,  &(2\,\me+\lle)\mLc&>0, &(2\,\me+\lle)(2\,\mh+\lh)\mLc&>0,
\end{align}
or analogously:
\begin{align}
2\,\me+\lle&>0,  &\mLc&>0, &2\,\mh+\lh&>0,
\end{align}
while for $k\rightarrow0$ we must have:
\begin{align}
2\,\me+\lle&>0,  &(4\,\mm+3\,\ke)(\me+\mh)&>0, &(\ke+\kh)(\me+\mh)(2\,\mm+\lm)&>0.
\end{align}
Since from \eqref{NecTran} we have necessarily $\me>0$, $\mh>0$, and from \eqref{NecSemi} we get $\ke+\kh\geq0$ and considering together the two limits for $k$ we obtain the necessary condition:
\begin{align}
2\,\me+\lle&>0, &2\,\mh+\lh&>0, &4\,\mm+3\,\ke&>0,&\ke+\kh&>0,\label{Nec1}\\\nonumber
\me&>0,&\mh&>0,&\mc&\geq0,&2\,\mm+\lm&>0.
\end{align}
Inspection shows that \eqref{Nec1} is our proposed sufficient condition \eqref{suff}. From $\me>0$ and $\mh>0$, it follows that $\mm>0$. Therefore condition \eqref{Nec1} is  \textbf{necessary} and \textbf{sufficient}. We have shown our main proposition:
\begin{proposition} The dynamic relaxed micromorphic model (eq.$\,$\eqref{eq:Dyn}) admits real planar waves if and only if 
\begin{align}
\mc&\geq0,&
\me&>0,&
2\,\me+\lle&>0,
\label{Prop}\\\nonumber
&&
\mh&>0,
 &\qquad
 2\,\mh+\lh&>0,\\\nonumber 
 &&
(\mm&>0),&
2\,\mm+\lm&>0,
 \qquad
\\\nonumber
\hspace{2cm}&\hspace{2.5cm}&
\ke+\kh&>0,
 &\hspace{2cm} 4\,\mm+3\,\ke&>0.&&\hspace{2.3cm}\blacksquare
\end{align}
\end{proposition}
 
 \hspace{-0.55cm}In \eqref{Prop} the requirement $\mm>0$ is redundant, since it is already assumed that $\me,\mh>0$. It is clear that positive definiteness of the elastic energy \eqref{DefPos} implies \eqref{Prop}. We remark that, as shown in the appendix \ref{Ineq}, the set of inequalities \eqref{Prop} is already implied by:
 \begin{empheq}[box=\widefbox]{align}
 \me>0,\qquad\mh>0,\qquad \mc\geq0,\qquad \ke+\kh>0,\qquad 2\,\mm+\lm>0.
 \end{empheq}	
Letting finally $\mh\rightarrow+\infty$ and $\kh\rightarrow+\infty$ (or $\mh\rightarrow+\infty$ and $\lh>\mathrm{const}$.) generates the limit condition for real wave velocities ($\me\rightarrow\mm$)
\begin{align}
\mm>0,\qquad \mc\geq0,\qquad 2\,\mm+\lm>0.\label{Macro}
\end{align}
which coincides, up to $\mc$, with the strong ellipticity condition in isotropic linear elasticity, see section \ref{ClassLin}, and it coincides fully with the condition for real wave velocities in micropolar elasticity, see section \ref{Cosserat}. A condition similar to \eqref{Macro} can be found in \cite[eq.$\,$8.14 p.$\,$ 26]{mindlin1963microstructure} where Mindlin requires that $\mm>0,\  2\,\mm+\lm>0$\footnote{Mindlin explains that such parameters ``are less than those that would be calculated from the strain-stiffnesses [of the unit cell]. This phenomenon is due to the compliance of the unit cell and has been found in a theory of crystal lattices by Gazis and Wallis \cite{gazis1962extensional}''.} (in our notation) which are obtained from the requirement of positive \textbf{group velocity} at $k=0$
\begin{align}
\frac{d \omega_\text{acoustic, long} (0)}{dk}>0,\qquad\qquad\frac{d \omega_\text{acoustic, trans} (0)}{dk}>0.
\end{align}

Let us emphasize that our method is not easily generalized to two immediate extensions. First, one could be interested in the isotropic relaxed micromorphic model with weighted inertia contributions and weighted curvatures \cite{dagostino2016panorama}. Second, one could be interested in the anisotropic setting \cite{barbagallo2016transparent}. In both cases the block-structure of the problem will be lost and one has to deal with the full $12\times12$ case, see equation \eqref{sys12} in the Appendix. Nonetheless, we expect positive-definiteness to always imply real wave propagation.

\begin{figure}[!h]
	\begin{centering}
		\begin{tabular}{ccccc}
			\includegraphics[width=5cm]{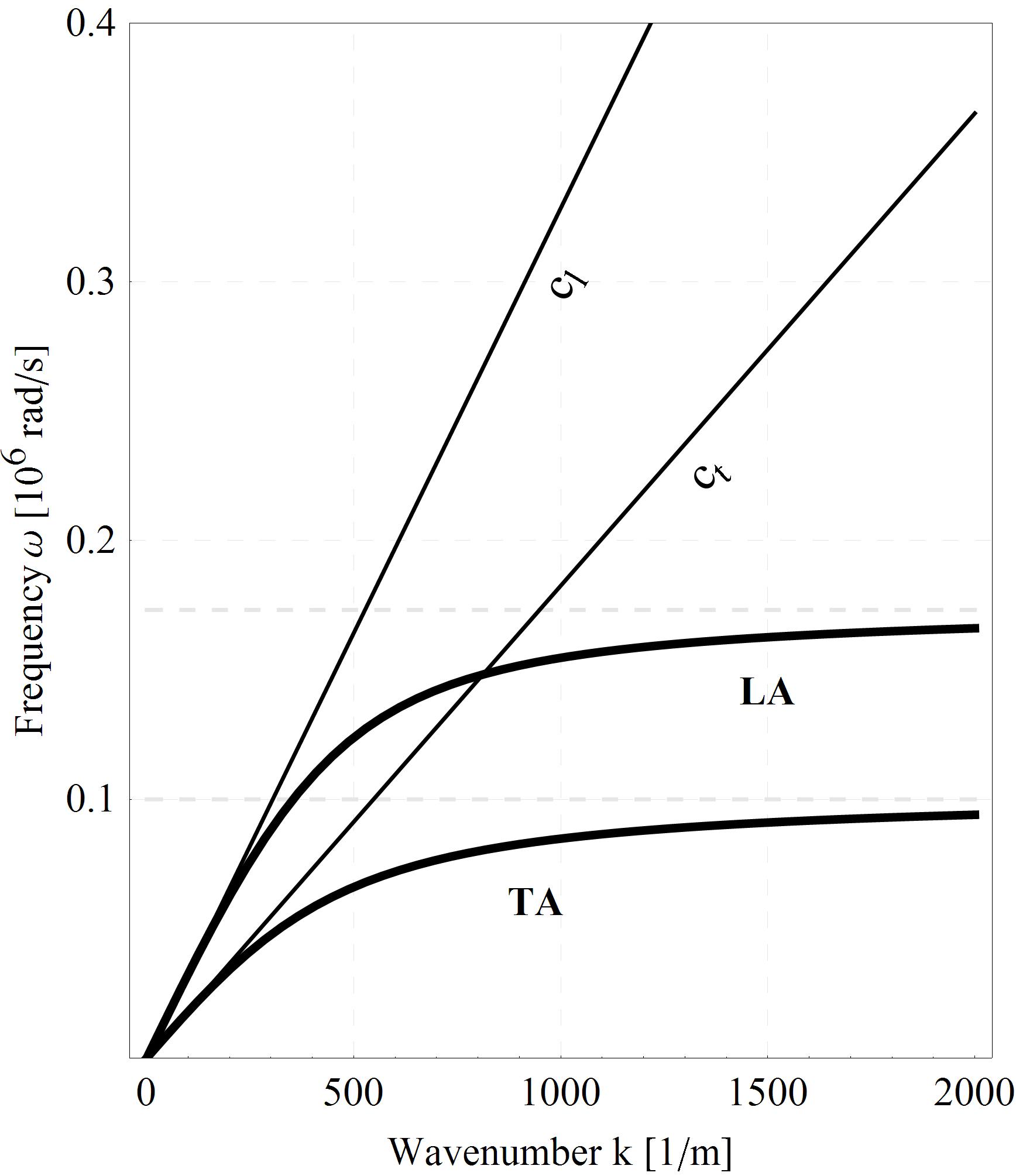}  &
			\includegraphics[width=5cm]{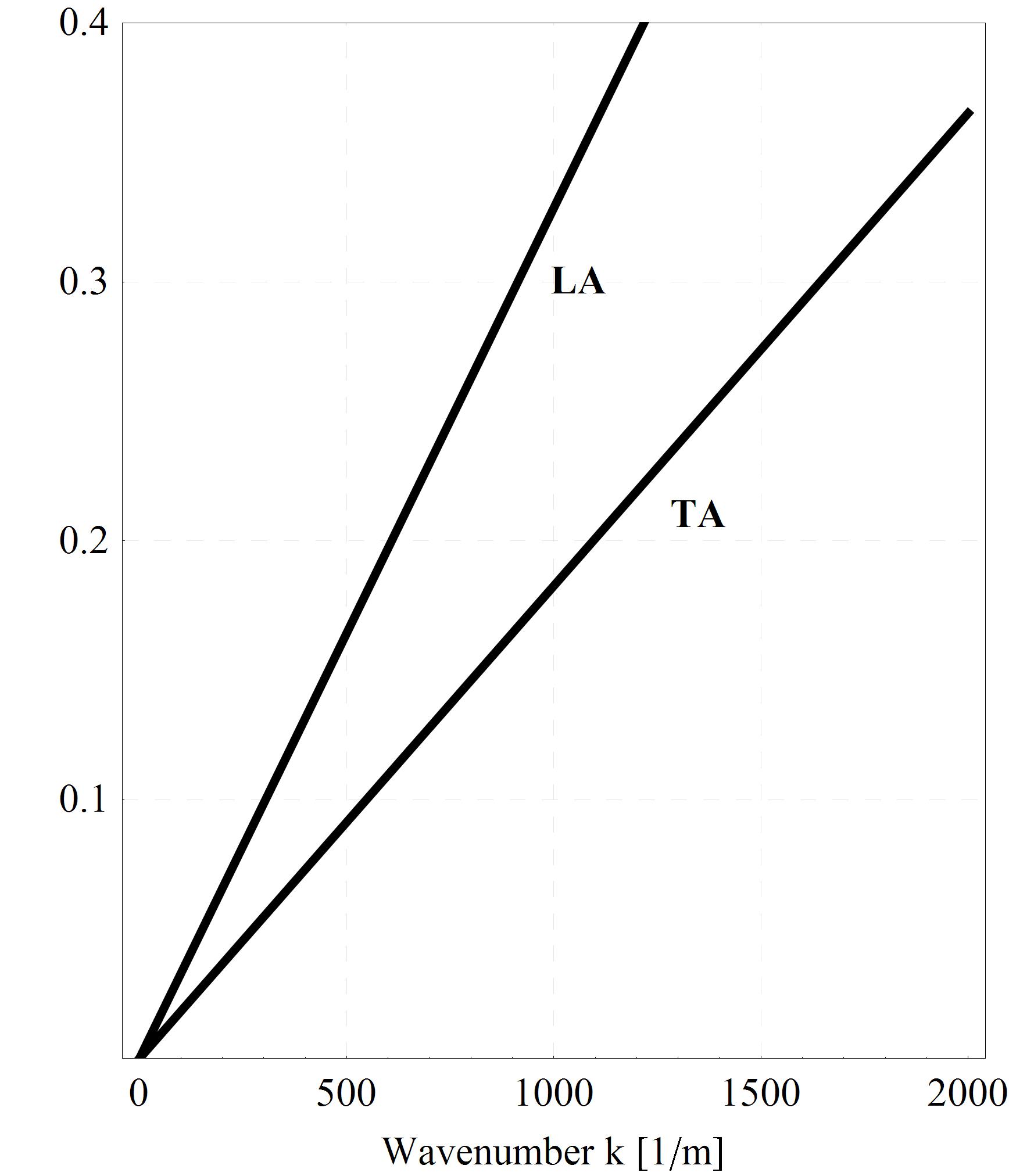}\\
			\quad(a)&\quad(b)
		\end{tabular}
		\par\end{centering}
	
	\caption{\label{Acoust}Dispersion relations $\omega=\omega(k)$ for the longitudinal acoustic wave LA, and the transverse acoustic TA in the
		\textbf{relaxed micromorphic model} (a) and in a classical Cauchy medium (b).
	}	
\end{figure}

In \cite{dagostino2016panorama} we show that the tangents of the acoustic branches in $k=0$ in the dispersion curves are
\begin{align}
c_l=\frac{d \omega_\text{acoustic, long} (0)}{dk}=	\sqrt{\frac{2\,\mm+\lm}{\rho}},\qquad \qquad\qquad c_t=\frac{d \omega_\text{acoustic, trans} (0)}{dk}=\sqrt{\frac{\mm}{\rho}}.
\end{align}
The tangents coincide with the classical linear elastic response if the latter has Lamé constants $\mm$ and $\lm$, as it is shown in Figure \ref{Acoust}.

\section{A comparison: classical isotropic linear elasticity \label{ClassLin}}
For classical linear elasticity with isotropic energy and kinetic energy:
\begin{align}
W(\nablau)=\mm\,\lVert \sym\left(\nablau\right)\rVert ^{2}+\frac{\lm}{2}\left(\mathrm{tr} \left(\nablau\right)\right)^{2},\qquad \qquad 	J=\frac{\rho}{2}\left\Vert u_{,t}\right\Vert ^{2}.\label{Energy1G}
\end{align}
The positive definiteness of the energy is equivalent to:
\begin{align}
\mm>0,\qquad\qquad\qquad 2\,\mm +3\, \lm>0.\label{cond1g}
\end{align}
It is easy to see that our homogenization formula \eqref{eq:IsotropicRel-2} implies \eqref{cond1g} under condition of positive definiteness of the relaxed micromorphic model.

The dynamical formulation is obtained  defining a joint Hamiltonian and assuming stationary action.  The dynamical equilibrium equations are:
\begin{align}
\rho\,  u_{,tt}=&\,\Div\left[2\,\mm\, \sym\left(\nablau\right)+\lm\tr\left(\nablau\right)\mathds{1}\right] . \label{Cauchy}
\end{align}
As before, in our study of wave propagation in micromorphic media we limit ourselves to the case of plane waves traveling in an infinite domain. We suppose that the space dependence of all introduced kinematic fields are limited to a direction defined by a unit vector $\widetilde{\xi}\in\R^3$ which is the direction of propagation of the wave. Therefore, we look for solutions of \eqref{Cauchy} in the form:
	\begin{align}
	u(x,t)=\widehat{u} \, e^{i \left(k\langle \widetilde{\xi},\, x\rangle_{\R^3}-\,\omega \,t\right)}\,,\qquad \widehat{u}\in\mathbb{C}^{3}\,,\quad \lVert\widetilde{\xi}\rVert^2=1\,.
	\end{align}
Since our formulation is isotropic, we can, without loss of generality, specify the direction $\widetilde{\xi}=e_1$. Then  $X=\langle e_1,x\rangle=x_1$, and we obtain:
	\begin{align}
	u(x,t)=\widehat{u} \, e^{i \left(k\, X-\,\omega \,t\right)}\,,\qquad \widehat{u}\in\mathbb{C}^{3}\,.
	\end{align}
	With this ansatz it is possible to write \eqref{Cauchy} as:
	\begin{align}
	\mathbf{A}_{5}(e_1,\omega,k) \, \widehat{u} =0\iff (	\mathcal{B}(e_1,k)-\omega^2 \id) \, \widehat{u} =0\,,
	\end{align}	
	where:
	\begin{align}
\mathbf{A}_{5}(e_1,\omega,k)  &=\Bigg(\begin{array}{ccc}	\frac{2\,\mm+\lm}{\rho} k^2 -\omega^2 & 0 	& 0 	\\
0 & \frac{\mm}{\rho} k^2-\omega^2  & 0 	\\ 
0&  0 	&\frac{\mm}{\rho}\, k^2-\omega^2 
	\end{array}	\Bigg),
	\\
	\mathcal{B}(e_1,k) &= \frac{k^2}{\rho}\,\Bigg(\begin{array}{ccc} 2\,\mm+\lm & 0 	& 0 	\\
	0 & \mm & 0 	\\ 
	0&  0 	& \mm
	\end{array}	\Bigg).
	\end{align}

Here, we observe that $\mathbf{A}_{5}(e_1,\omega,k)$ is already diagonal and real. Requesting real wave velocities means $\omega^2\geq0$. For $k\neq 0$, this leads to the classical so-called \textbf{strong ellipticity condition}:
\begin{align}
	\mm>0,\qquad\qquad 2\,\mm +\lm>0,\label{pos1g}
\end{align}
which is implied by positive  definiteness of the energy \eqref{cond1g}.

In classical (linear or nonlinear) elasticity, the condition of real wave propagation \eqref{pos1g} is equivalent to \textbf{strong ellipticity} and \textbf{rank-one convexity}. Indeed, rank-one convexity amounts to set ($\xi=k \widetilde{\xi}$ with $\lVert\xi\rVert^2=1$):
\begin{align}
\frac{d^2}{dt^2}\bigg\rvert_{t=0} \ W(\nablau+t\, \widehat{u} \,\otimes \xi)\geq 0 \iff \langle \Csym \left(\widehat{u} \,\otimes \xi\right),\widehat{u} \,\otimes \xi \rangle \geq 0, \label{Rank1}
\end{align}
where $\Csym$ is the fourth-order elasticity tensor. Condition \eqref{Rank1} reads then:
\begin{align}\nonumber
	0&\le 2\, \mm\,\lVert \sym\left(\widehat{u} \,\otimes \xi\right)\rVert ^{2}+\lm\left(\mathrm{tr} \left(\widehat{u} \,\otimes \xi\right)\right)^{2}
\label{Ellipt}
	= 	\mm\, \lVert\widehat{u}\rVert^2 \lVert\xi\rVert^2+( \mm +\lm )\langle \widehat{u} , \xi \rangle^2.
\end{align}
We may express \eqref{Ellipt} given $\xi\in\R^3$ as a quadratic form in $\widehat{u}\in\R^3$, which results in:
\begin{align}
\mm\, \lVert\widehat{u}\rVert^2 \lVert\xi\rVert^2+( \mm +\lm )\langle \widehat{u} , \xi \rangle^2
=\langle\mathcal{D}(\xi) \widehat{u},\widehat{u}\rangle,
\end{align} 
where the components of the symmetric and real $3\times3$ matrix $\mathcal{D}(\xi)$ read
\begin{align}
\mathcal{D}(\xi)=\,&\Bigg(\begin{array}{ccc
	}
 \hspace{-0.1cm}(2\,\mm+\lm)\xi_1^2+\mm(\xi_2^2+\xi_3^2) & (\lm+\mm) \xi_1\,\xi_2
\\
 (\lm+\mm) \xi_1\,\xi_2 & \hspace{-0.25cm}(2\,\mm+\lm)\xi_2^2 +\mm(\xi_1^2+\xi_3^2) 
  \\
 (\lm+\mm) \xi_1\,\xi_3&  (\lm+\mm) \xi_1\xi_2 
\end{array}
\\&\quad
\begin{array}{c} (\lm+\mm) \xi_1\,\xi_3\\(\lm+\mm) \xi_2\,\xi_3 \\2\,\mm+\lm)\xi_3^2+\mm(\xi_1^2+\xi_2^2) \end{array}
\Bigg).\nonumber
\end{align}
The three principal invariants are independent of the direction $\xi$ due to isotropy and are given by:
\begin{align}
\tr (\mathcal{D}(\xi))&=\lVert\xi\rVert^2 (4\, \mm+\lm)=k^2(4 \,\mm+\lm), \nonumber\\
\tr (\mathrm{Cof}\,\mathcal{D}(\xi))&=\lVert\xi\rVert^4 \mm (5\,\mm+2\,\lm)= 
k^4 \mm (5 \,\mm+2\,\lm), \\
\det (\mathcal{D}(\xi))&=\lVert\xi\rVert^6\mm^2 (2, \mm+\lm)=
k^6 \mm^2 (2\, \mm+\lm). \nonumber
\end{align}
Since $\mathcal{D}(\xi)$ is real and symmetric, its eigenvalues are real. The eigenvalues of the matrix $\mathcal{D}(\xi)$ are \linebreak$k^2(2\,\mm+\lm)$, $k^2 \mm$ and $k^2 \mm$ such that positivity at $k\neq0$ is satisfied if and only if\footnote{The eigenvalues of $\mathcal{D}(\xi)$ are independent of the propagation direction $\xi\in\R^3$ which makes sense for the isotropic formulation at hand.}:
\begin{align}
\mm>0,\qquad\qquad 2\,\mm + \lm>0,
\end{align}
which are the usual strong ellipticity conditions. We note here that the latter calculations also show that $\mathcal{B}(e_1)=\frac{1}{\rho}\,k^2\,\mathcal{D}(e_1)$. Alternatively, one may directly form the so-called \textbf{acoustic tensor} $B(\xi)\in\R^{3\times3}$ by
\begin{align}
 B(\xi).\widehat{u}:=[\Csym.(\widehat{u}\otimes\xi)].\xi,\quad\forall\widehat{u}\in\R^3,\label{acous}
 \end{align} 
 in indices we have $(B(\xi))_{ij}=\Csym^{ikjl} \widehat{u}_k \widehat{u}_l\neq \Csym.(\xi\otimes\xi)$. With \eqref{acous} we obtain\footnote{$[\Csym(\widehat{u}\otimes\xi)](\widehat{u}\otimes\xi)\neq\Csym[(\widehat{u}\otimes\xi)(\widehat{u}\otimes\xi)]$.}:
 \begin{align}
 \langle\widehat{u},B(\xi).\widehat{u}\rangle_{\R^3}&=\langle \underbrace{[\Csym.(\widehat{u}\otimes\xi)]}_{=:\widehat{B}\in\R^{3\times3}}\xi,\widehat{u}\rangle_{\R^3}=\langle\widehat{B}\,\xi,\widehat{u}\rangle_{\R^3}
 	=\langle\widehat{B}\,(\xi\otimes \widehat{u}),\mathds{1}\rangle_{\R^{3\times3}}=\langle\widehat{B},(\xi\otimes \widehat{u})^T\rangle_{\R^{3\times3}}
 \\&=\langle\widehat{B},\widehat{u}\otimes \xi\rangle_{\R^{3\times3}}=\langle\Csym\, (\widehat{u}\otimes \xi),\widehat{u}\otimes \xi\rangle_{\R^{3\times3}},\nonumber
 \end{align}
 and we see that strong ellipticity $\langle\Csym\, (\widehat{u}\otimes \xi),\widehat{u}\otimes \xi\rangle_{\R^{3\times3}}>0$ is equivalent to the positive definiteness of the acoustic tensor $B(\xi)$. 


\section{A further comparison: the Cosserat model\label{Cosserat}}
In the isotropic hyperelastic case the elastic energy density and the kinetic energy of the Cosserat model read:
	\begin{align}
	W=&\ \mm\,\lVert \sym\left(\nablau\right)\rVert ^{2}+\mc\,\lVert \skew\left(\nablau-A\right)\rVert ^{2}+\frac{\lm}{2}\left(\mathrm{tr} \left(\nablau\right)\right)^{2}+\frac{\mm L_c^2}{2} \,\lVert \Curl A\rVert^2,\\\nonumber
	J=&\ \frac{\rho}{2}\left\Vert u_{,t}\right\Vert ^{2}+\frac{\eta}{2}\left\Vert A_{,t}\right\Vert ^{2},
	\end{align}
where $A\in\so(3)$, can be expressed as a function of $a\in\R^3$ as:
\begin{align}
A=\mathrm{anti}(a)=\left(\begin{array}{ccc}
0& -a_{3} &a_{2}\\
a_{3}& 0 & -a_{1}\\
-a_{2}&a_{1}&0
\end{array}\right).
\end{align}
Here, we assume for clarity a uni-constant curvature expression in terms of only $\lVert\Curl\,A\rVert^2$. 	Strict positive definiteness of the potential energy is equivalent to the following simple relations for the introduced parameters
\begin{align}
2\,\mm+3\,\lm>0,\qquad\qquad \mm >0, \qquad\qquad \mc>0,\qquad\qquad L_c>0. \label{posCoss}
\end{align}
The dynamical formulation is obtained  defining a joint Hamiltonian and assuming stationary action.  The dynamical equilibrium equations are:
\begin{align}
\rho\,  u_{,tt}=&\,\Div\left[2\,\mm\, \sym\left(\nablau-A\right)+2\,\mc \,\skew\left(\nablau-A\right)+\lm\tr\left(\nablau-A\right)\mathds{1}\right] , \nonumber \\
\eta  A_{,tt}=&\,-\mm L_c^2 \Curl \Curl A+2 \,\mc\, \skew\left(\nablau-A\right),\nonumber 
\end{align}
see also \cite{jeong2008existence,jeong2009numerical,neff2009new,neff2010stable} for formulations in terms of axial vectors. Considering plane and stationary waves of amplitudes $\widehat{u}$ and $\widehat{a}$, it is possible to express this system as:
\begin{align}
\mathbf{A}_6 (\omega,k) \cdot \left(\begin{array}{cc} \widehat{u}_1& \widehat{a}_{1}
\end{array}\right)^T&=0,&\mathbf{A}_7 (\omega,k) \cdot \left(\begin{array}{cc} \widehat{u}_2& -\widehat{a}_{3}
\end{array}\right)^T&=0,&\mathbf{A}_7 (\omega,k) \cdot \left(\begin{array}{cc} \widehat{u}_3& \widehat{a}_{2}
\end{array}\right)^T&=0,
\end{align}
where
\begin{align}
\mathbf{A}_6 (\omega,k)&=\left(\begin{array}{cc} k^2 (2\,\mm+\lm))/\rho -\omega^2& 0\\ 0& (2\mm L_c^2 k^2+2\mc)/\eta-\omega^2\end{array}\right),\\\nonumber \\
\mathbf{A}_7 (\omega,k)&=\left(\begin{array}{cc} 
k^2(\mm+\mc)/\rho -\omega^2 &-2 i k\mc /\rho \\
i k \mc/\eta & (k^2 \mm L_c^2+ 4\mc)/(2\eta)-\omega^2
\end{array}\right).
\end{align}
As done in the case of the relaxed micromorphic model, it is possible to express equivalently the problem with $\mathbf{A}_6 (\omega,k)$ and the following symmetric matrix:
\begin{align}
\mathbf{\overline{A}}_7 (k)&=\mathrm{diag}_7 \cdot \mathbf{A}_7 (\omega,k)\cdot \mathrm{diag}_7^{-1}=\left(\begin{array}{cc}
k^2(\mm+\mc)/\rho-\omega^2  &\sqrt{2}  k\mc /\sqrt{\rho\eta} \\
\sqrt{2}  k\mc /\sqrt{\rho\eta} & (k^2 \mm L_c^2+ 4\mc)/(2\eta)-\omega^2
\end{array}\right),
\end{align}
where
\begin{align}
\mathrm{diag}_7=
\left(\begin{array}{cc}\sqrt{\rho}&0\\0&i \sqrt{2\eta} \end{array}\right).\label{eigCoss}
\end{align}
Since $\omega^2$ appears only on the diagonal, the problem can be analogously expressed as the following eigenvalue-problems:
\begin{equation}
\mathrm{det}\,\left(\mathbf{B}_{6}(k)-\omega^2\,\mathds{1}\right)=0,\qquad\qquad\qquad \mathrm{det}\,\left(\mathbf{B}_{7}(k)-\omega^2\,\mathds{1}\right)=0,
\end{equation}
where
\begin{align}
\mathbf{B}_6 (k)&=\left(\begin{array}{cc} k^2 (2\,\mm+\lm))/\rho & 0\\ 0& (2\mm L_c^2 k^2+2\mc)/\eta^2\end{array}\right),\\\nonumber\\\mathbf{B}_7 (k)&=\left(\begin{array}{cc}
k^2(\mm+\mc)/\rho  &\sqrt{2}  k\mc /\sqrt{\rho\eta} \\
\sqrt{2}  k\mc /\sqrt{\rho\eta} & (k^2 \mm L_c^2+ 4\mc)/(2\eta)
\end{array}\right),
\end{align}
are the blocks of the acoustic tensor $\mathbf{B}$
\begin{align}
\mathbf{B}(k)&=\left(\begin{array}{ccc}  \mathbf{B}_6 & \mathbf{0}&\mathbf{0}\\ \mathbf{0}& \mathbf{B}_7&\mathbf{0}\\\mathbf{0}& \mathbf{0}& \mathbf{B}_7\end{array}\right).
\end{align} 
The eigenvalues of the matrix $\mathbf{B}_6 (k)$ are simply the elements of the diagonal, therefore we have:
\begin{align}
\omega_\text{acoustic, long}(k)=k\, \sqrt{ \frac{ 2\,\mm+\lm}{\rho}}, \qquad \qquad \omega_\text{optic, long}(k)=\sqrt{\frac{2\mm L_c^2 k^2+2\mc}{\eta}},
\end{align}
while for $\mathbf{B}_7 (k)$ it is possible to find:
\begin{align}
\omega_\text{acoustic, trans}(k)=\sqrt{a(k)-\sqrt{a(k)^2-b^2 k^2}}, \qquad \qquad \omega_\text{optic, trans}(k)=\sqrt{a(k)+\sqrt{a(k)^2-b^2 k^2}},
\end{align}
where we have set:
\begin{align}
a(k)=\frac{4\, \mc + \mm L_c^2 k^2}{\eta}  + 
 2\,\frac{\mm + \mc}{\rho}\,k^2, \qquad \qquad b^2= 8\, \frac{\mm (4 \mc + k^2 L_c^2 (\mm + \mc))}{ \rho \, \eta }.
\end{align}
The acoustic branches are those curves $\omega=\omega(k)$ as solutions of \eqref{eigCoss} that satisfy $\omega(0)=0$. We note here that the acoustic branches of the longitudinal and transverse dispersion curves have as tangent in $k=0$\footnote{To obtain the slopes in 0 it is possible to search for a solution of the type $\omega=a\,k$ and then evaluate the limit for $a\rightarrow0$, see \cite{dagostino2016panorama} for a thorough explanation in the relaxed micromorphic case.}
\begin{align}
c_l=\frac{d \omega_\text{acoustic, long} (0)}{dk}=	\sqrt{\frac{2\,\mm+\lm}{\rho}},\qquad \qquad\qquad c_t=\frac{d \omega_\text{acoustic, trans} (0)}{dk}=\sqrt{\frac{\mm}{\rho}}, \label{CossTang}
\end{align}
respectively. Moreover, the longitudinal acoustic branch is non-dispersive, i.e. a straight line with slope \eqref{CossTang}$_1$.
The matrix $\mathbf{B}_6 (k)$  is positive-definite for arbitrary $k\neq0$ if:
\begin{align}
2\,\mm+\lm>0,\qquad \mm>0, \qquad \mc\geq0, \label{Coss1} 
\end{align}
Using the Sylvester criterion, $\mathbf{B}_7 (k)$  is positive-definite if and only if the principal minors are positive, namely:
\begin{align}
(\mathbf{B}_7)_{11}&
=k^2\,\frac{(\mm+\mc)}{\rho}>0,\\\nonumber
\det(\mathbf{B}_7)&
=\frac{k^2}{2\eta \rho} (4 \mm\, \mc + k^2 \mm L_c^2 (\mm + \mc))>0,
\end{align}
from which we obtain the condition:
\begin{align}
\mm+\mc>0,\qquad \mm>0, \qquad \mc\geq0. \label{Coss2}
\end{align}
Considering these two sets of conditions, it is possible to state a \textbf{necessary}  and \textbf{sufficient} condition for the positive definiteness of $\mathbf{B}_6 (k)$ and $\mathbf{B}_7 (k)$ and therefore of the acoustic tensor $\mathbf{B}(k)$:
\begin{align}
2\,\mm+\lm>0,\qquad\qquad \mm >0, \qquad\qquad \mc\geq0. \label{condCoss}
\end{align}
which are implied by the positive-definiteness of the energy \eqref{posCoss}. Eringen \cite[p.150]{eringen1999microcontinuum} also obtains correctly \eqref{Coss1} \textbf{and} \eqref{Coss2} (in his notation $\mc=\kappa/2$, $\mm=\mu_{\mathrm{Eringen}}+\kappa/2$).

 
In \cite{altenbach2010acceleration,eremeyev2005acceleration} strong ellipticity for the Cosserat-micropolar model is defined and investigated. 
In this respect we note that ellipticity is connected to acceleration waves while our investigation concerns real wave velocities for planar waves. Similarly to \cite{neff2008relations} it is established in \cite{altenbach2010acceleration,eremeyev2005acceleration} that strong ellipticity for the micropolar model holds if and only if (the uni-constant curvature case in our notation):
\begin{align}
2\,\mm+\lm>0,\qquad\qquad \mm+\mc>0. \label{ellipt}
\end{align}
We conclude that for micropolar material models, (and therefore also for micromorphic materials) strong ellipticity \eqref{ellipt} is too weak to ensure real planar waves since it is implied by, but does not imply \eqref{condCoss}. This fact seems not to have been well appreciated before.

\section{Acknowledgments}
We thank Victor A. Eremeyev for helpful clarification. The work of Ionel-Dumitrel Ghiba was supported by a grant of the Romanian National Authority for Scientific Research and Innovation, CNCS-UEFISCDI, project number PN-II-RU-TE-2014-4-1109. Angela Madeo thanks INSA-Lyon for the funding of the BQR 2016 \textquotedbl{}Caractérisation mécanique inverse des métamatériaux: modélisation, identification expérimentale des paramètres et évolutions possibles\textquotedbl{}.

\footnotesize

\let\stdsection\section
\def\section*#1{\stdsection{#1}}

\bibliography{library}

\begin{thebibliography}{10}

\bibitem{abreu2016refined}
Rafael Abreu, Jochen Kamm, Angela Madeo, and Patrizio Neff.
\newblock {Refined modelling of rotational waves in seismology with the relaxed
  micromorphic model}.
\newblock {\em In preparation}, 2016.

\bibitem{altenbach2010acceleration}
Holm Altenbach, Victor~A. Eremeyev, Leonid~P. Lebedev, and Leonardo~A.
  Rend{\'{o}}n.
\newblock {Acceleration waves and ellipticity in thermoelastic micropolar
  media}.
\newblock {\em Archive of Applied Mechanics}, 80(3):217--227, 2010.

\bibitem{balzani2006polyconvex}
Daniel Balzani, Patrizio Neff, J{\"{o}}rg Schr{\"{o}}der, and Gerhard~A.
  Holzapfel.
\newblock {A polyconvex framework for soft biological tissues. Adjustment to
  experimental data}.
\newblock {\em International Journal of Solids and Structures},
  43(20):6052--6070, 2006.

\bibitem{barbagallo2016transparent}
Gabriele Barbagallo, Marco~Valerio {d}'Agostino, Rafael Abreu, Ionel-Dumitrel
  Ghiba, Angela Madeo, and Patrizio Neff.
\newblock {Transparent anisotropy for the relaxed micromorphic model:
  macroscopic consistency conditions and long wave length asymptotics}.
\newblock {\em Preprint ArXiv}, 1601.03667, 2016.

\bibitem{bauer2014new}
Sebastian Bauer, Patrizio Neff, Dirk Pauly, and Gerhard Starke.
\newblock {New Poincar{\'{e}}-type inequalities}.
\newblock {\em Comptes Rendus Mathematique}, 352(2):163--166, 2014.

\bibitem{bauer2016dev}
Sebastian Bauer, Patrizio Neff, Dirk Pauly, and Gerhard Starke.
\newblock {Dev-Div- and DevSym-DevCurl-inequalities for incompatible square
  tensor fields with mixed boundary conditions}.
\newblock {\em ESAIM: Control, Optimisation and Calculus of Variations},
  22(1):112--133, 2016.

\bibitem{chirita2007strong}
Stan Chiri\c{t}\u{a}, Alexandre Danescu, and Michele Ciarletta.
\newblock {On the strong ellipticity of the anisotropic linearly elastic
  materials}.
\newblock {\em Journal of Elasticity}, 87(1):1--27, 2007.

\bibitem{chirita2009strong}
Stan Chiri\c{t}\u{a} and Ionel-Dumitrel Ghiba.
\newblock {Strong ellipticity and progressive waves in elastic materials with
  voids}.
\newblock {\em Proceedings of the Royal Society A: Mathematical, Physical and
  Engineering Sciences}, 466(2114):439--458, 2009.

\bibitem{dagostino2016panorama}
Marco~Valerio {d}'Agostino, Gabriele Barbagallo, Ionel-Dumitrel Ghiba, Rafael
  Abreu, Angela Madeo, and Patrizio Neff.
\newblock {A panorama of dispersion curves for the isotropic weighted relaxed
  micromorphic model}.
\newblock {\em In preparation}, 2016.

\bibitem{eremeyev2005acceleration}
Victor~A. Eremeyev.
\newblock {Acceleration waves in micropolar elastic media}.
\newblock {\em Doklady Physics}, 50(4):204--206, 2005.

\bibitem{eringen1999microcontinuum}
Ahmed~Cemal Eringen.
\newblock {\em {Microcontinuum field theories}}.
\newblock Springer-Verlag, New York, 1999.

\bibitem{gazis1962extensional}
Denos~C. Gazis and Richard~F. Wallis.
\newblock {Extensional waves in cubic crystals plates}.
\newblock In {\em Proceedings of the 4th U.S. National Congress in Applied
  Mechanics}, pages 161--168, 1962.

\bibitem{ghiba2014relaxed}
Ionel-Dumitrel Ghiba, Patrizio Neff, Angela Madeo, Luca Placidi, and Giuseppe
  Rosi.
\newblock {The relaxed linear micromorphic continuum: existence, uniqueness and
  continuous dependence in dynamics}.
\newblock {\em Mathematics and Mechanics of Solids}, 20(10):1171--1197, 2014.

\bibitem{gilbert1991positive}
George~T. Gilbert.
\newblock {Positive definite matrices and Sylvester's criterion}.
\newblock {\em The American Mathematical Monthly}, 98(1):44--46, 1991.

\bibitem{greene2012microelastic}
Steven Greene, Stefano Gonella, and Wing~Kam Liu.
\newblock {Microelastic wave field signatures and their implications for
  microstructure identification}.
\newblock {\em International Journal of Solids and Structures},
  49(22):3148--3157, 2012.

\bibitem{jeong2008existence}
Jena Jeong and Patrizio Neff.
\newblock {Existence, uniqueness and stability in linear Cosserat elasticity
  for weakest curvature conditions}.
\newblock {\em Mathematics and Mechanics of Solids}, 15(1):78--95, 2010.

\bibitem{jeong2009numerical}
Jena Jeong, Hamidr{\'{e}}za Ram{\'{e}}zani, Ingo M{\"{u}}nch, and Patrizio
  Neff.
\newblock {A numerical study for linear isotropic Cosserat elasticity with
  conformally invariant curvature}.
\newblock {\em Zeitschrift f{\"{u}}r Angewandte Mathematik und Mechanik},
  89(7):552--569, 2009.

\bibitem{madeo2016first}
Angela Madeo, Gabriele Barbagallo, Marco~Valerio {d}'Agostino, Luca Placidi,
  and Patrizio Neff.
\newblock {First evidence of non-locality in real band-gap metamaterials:
  determining parameters in the relaxed micromorphic model}.
\newblock {\em to appear in Proceedings of the Royal Society A: Mathematical,
  Physical and Engineering Sciences}, 2016.

\bibitem{madeo2014band}
Angela Madeo, Patrizio Neff, Ionel-Dumitrel Ghiba, Luca Placidi, and Giuseppe
  Rosi.
\newblock {Band gaps in the relaxed linear micromorphic continuum}.
\newblock {\em Zeitschrift f{\"{u}}r Angewandte Mathematik und Mechanik},
  95(9):880--887, 2014.

\bibitem{madeo2015wave}
Angela Madeo, Patrizio Neff, Ionel-Dumitrel Ghiba, Luca Placidi, and Giuseppe
  Rosi.
\newblock {Wave propagation in relaxed micromorphic continua: modeling
  metamaterials with frequency band-gaps}.
\newblock {\em Continuum Mechanics and Thermodynamics}, 27(4-5):551--570, 2015.

\bibitem{madeo2016reflection}
Angela Madeo, Patrizio Neff, Ionel-Dumitrel Ghiba, and Giussepe Rosi.
\newblock {Reflection and transmission of elastic waves in non-local band-gap
  metamaterials: a comprehensive study via the relaxed micromorphic model}.
\newblock {\em Journal of the Mechanics and Physics of Solids}, 2016.

\bibitem{merodio2006note}
Jose Merodio and Patrizio Neff.
\newblock {A note on tensile instabilities and loss of ellipticity for a
  fiber-reinforced nonlinearly elastic solid}.
\newblock {\em Archives of Mechanics}, 58(3):293--303, 2006.

\bibitem{mindlin1963microstructure}
Raymond~David Mindlin.
\newblock {Microstructure in linear elasticity}.
\newblock Technical report, Office of Naval Research, 1963.

\bibitem{mindlin1964micro}
Raymond~David Mindlin.
\newblock {Micro-structure in linear elasticity}.
\newblock {\em Archive for Rational Mechanics and Analysis}, 16(1):51--78,
  1964.

\bibitem{munch2016rotational}
Ingo M{\"{u}}nch and Patrizio Neff.
\newblock {Rotational invariance conditions in elasticity, gradient elasticity
  and its connection to isotropy}.
\newblock {\em Preprint ArXiv}, 1603.06153, 2016.

\bibitem{neff2004material}
Patrizio Neff.
\newblock {On material constants for micromorphic continua}.
\newblock In {\em Trends in Applications of Mathematics to Mechanics, STAMM
  Proceedings, Seeheim}, pages 337--348. Shaker--Verlag, 2004.

\bibitem{neff2006cosserat}
Patrizio Neff.
\newblock {The Cosserat couple modulus for continuous solids is zero viz the
  linearized Cauchy-stress tensor is symmetric}.
\newblock {\em Zeitschrift f{\"{u}}r Angewandte Mathematik und Mechanik},
  86(11):892--912, 2006.

\bibitem{neff2008relations}
Patrizio Neff.
\newblock {Relations of constants for isotropic linear Cosserat elasticity
  ({http://www.uni-due.de/\%7ehm0014/Cosserat\_files/web\_coss\_relations.pdf},
  typos in equations 2.8,2.9,2.10)}.
\newblock Technical report, Fachbereich Mathematik, Technische Universituat
  Darmstadt, Darmstadt, Germany, 2008.

\bibitem{neff2007geometrically}
Patrizio Neff and Samuel Forest.
\newblock {A geometrically exact micromorphic model for elastic metallic foams
  accounting for affine microstructure. Modelling, existence of minimizers,
  identification of moduli and computational results}.
\newblock {\em Journal of Elasticity}, 87(2-3):239--276, 2007.

\bibitem{neff2015relaxed}
Patrizio Neff, Ionel-Dumitrel Ghiba, Markus Lazar, and Angela Madeo.
\newblock {The relaxed linear micromorphic continuum: well-posedness of the
  static problem and relations to the gauge theory of dislocations}.
\newblock {\em The Quarterly Journal of Mechanics and Applied Mathematics},
  68(1):53--84, 2014.

\bibitem{neff2014unifying}
Patrizio Neff, Ionel-Dumitrel Ghiba, Angela Madeo, Luca Placidi, and Giuseppe
  Rosi.
\newblock {A unifying perspective: the relaxed linear micromorphic continuum}.
\newblock {\em Continuum Mechanics and Thermodynamics}, 26(5):639--681, 2014.

\bibitem{neff2009new}
Patrizio Neff and Jena Jeong.
\newblock {A new paradigm: the linear isotropic Cosserat model with conformally
  invariant curvature energy}.
\newblock {\em Zeitschrift f{\"{u}}r Angewandte Mathematik und Mechanik},
  89(2):107--122, 2009.

\bibitem{neff2010stable}
Patrizio Neff, Jena Jeong, and Andreas Fischle.
\newblock {Stable identification of linear isotropic Cosserat parameters:
  bounded stiffness in bending and torsion implies conformal invariance of
  curvature}.
\newblock {\em Acta Mechanica}, 211(3-4):237--249, 2010.

\bibitem{neff2011canonical}
Patrizio Neff, Dirk Pauly, and Karl-Josef Witsch.
\newblock {A canonical extension of Korn's first inequality to H(Curl)
  motivated by gradient plasticity with plastic spin}.
\newblock {\em Comptes Rendus Mathematique}, 349(23):1251--1254, 2011.

\bibitem{neff2012maxwell}
Patrizio Neff, Dirk Pauly, and Karl-Josef Witsch.
\newblock {Maxwell meets Korn: A new coercive inequality for tensor fields in
  RNxN with square-integrable exterior derivative}.
\newblock {\em Mathematical Methods in the Applied Sciences}, 35(1):65--71,
  2012.

\bibitem{neff2015poincare}
Patrizio Neff, Dirk Pauly, and Karl-Josef Witsch.
\newblock {Poincar{\'{e}} meets Korn via Maxwell: Extending Korn's first
  inequality to incompatible tensor fields}.
\newblock {\em Journal of Differential Equations}, 258(4):1267--1302, 2015.

\bibitem{nowacky1985theory}
Witold Nowacki.
\newblock {Theory of asymmetric elasticity}, 1985.

\bibitem{romano2016micromorphic}
Giovanni Romano, Raffaele Barretta, and Marina Diaco.
\newblock {Micromorphic continua: non-redundant formulations}.
\newblock {\em Continuum Mechanics and Thermodynamics}, 2016.

\bibitem{schroder2002application}
J{\"{o}}rg Schr{\"{o}}der and Patrizio Neff.
\newblock {Application of polyconvex anisotropic free energies to soft
  tissues}.
\newblock {\em 5th World Congress on Computational Mechanics}, 2002.

\bibitem{schroder2008anisotropic}
J{\"{o}}rg Schr{\"{o}}der, Patrizio Neff, and Vera Ebbing.
\newblock {Anisotropic polyconvex energies on the basis of crystallographic
  motivated structural tensors}.
\newblock {\em Journal of the Mechanics and Physics of Solids},
  56(12):3486--3506, 2008.

\bibitem{smith1967waves}
A.C. Smith.
\newblock {Waves in micropolar elastic solids}.
\newblock {\em International Journal of Engineering Science}, 5(10):741--746,
  1967.

\bibitem{smith1968inequalities}
A.C. Smith.
\newblock {Inequalities between the constants of a linear micro-elastic solid}.
\newblock {\em International Journal of Engineering Science}, 6(2):65--74,
  1968.

\end{thebibliography}
\bibliographystyle{plain}

\let\section\stdsection

\section{Appendix}
\subsection{Inequality relations between material parameters \label{Ineq}}

The formulas in section \ref{EN} are based on the harmonic mean of two numbers $\ke$ and $\kh$ (or $\me$ and $\mh$). If the two numbers are positive, it is easy to see that:
\begin{align}
\km\leq \mathrm{min}(\ke,\kh).
\end{align}
Here, we show that the same conclusion still holds if we merely assume that $\ke+\kh>0$. This allows for either $\ke<0$ or $\kh<0$.
Therefore, considering that $ \ke+\kh>0$, even if the energy is not strictly positive, it is possible to derive that:
\begin{align}
\km=\frac{\kh\,\ke}{\ke+\kh}=\frac{\kh\,\ke+\ke^2-\ke^2}{\ke+\kh}=\ke\frac{\kh+\ke}{\ke+\kh}-\frac{\ke^2}{\ke+\kh}=\ke\underbrace{-\frac{\ke^2}{\ke+\kh}}_{\leq 0}&\leq\ke,\label{dis1}\\\nonumber
\km=\frac{\kh\,\ke}{\ke+\kh}=\frac{\kh\,\ke+\kh^2-\kh^2}{\ke+\kh}=\kh\frac{\kh+\ke}{\ke+\kh}-\frac{\kh^2}{\ke+\kh}=\kh\underbrace{-\frac{\kh^2}{\ke+\kh}}_{\leq 0}&\leq\kh.
\end{align}
Considering similarly $\me+\mh>0$, it is possible to obtain:
\begin{align}
\mm=\frac{\mh\,\me}{\me+\mh}=\frac{\mh\,\me+\me^2-\me^2}{\me+\mh}=\me\frac{\mh+\me}{\me+\mh}-\frac{\me^2}{\me+\mh}=\me\underbrace{-\frac{\me^2}{\me+\mh}}_{\leq 0}&\leq\me,\label{dis2}\\\nonumber
\mm=\frac{\mh\,\me}{\me+\mh}=\frac{\mh\,\me+\mh^2-\mh^2}{\me+\mh}=\mh\frac{\mh+\me}{\me+\mh}-\frac{\mh^2}{\me+\mh}=\mh\underbrace{-\frac{\mh^2}{\me+\mh}}_{\leq 0}&\leq\mh.
\end{align}
Therefore, if $\me+\mh>0$ and $\ke+\kh>0$, the macroscopic parameters are less or equal than respective microscopic parameters, namely:
\begin{align}
\ke\geq\km,\qquad\qquad\kh\geq\km\qquad\qquad \me\geq\mm,\qquad\qquad \mh\geq\mm,
\end{align} 
and it is possible to show that:
\begin{align}
2\,\me+\lle&=
\frac{1}{3}(4\,\me+3\,\ke )\geq\frac{1}{3}(4\,\mm+3\,\km)=2\,\mm+\lm>0,\nonumber\\
2\,\mh+\lh&=
\frac{1}{3}(4\,\mh+ 3\,\kh)\geq\frac{1}{3}( 4\,\mm+3\,\km)
=2\,\mm+\lm>0,\\
(2\,\me+\lle)+(2\,\mh+\lh)&\geq 2\,(2\,\mm+\lm)>0,\nonumber\\
4\,\mm+3\,\ke&\geq  4\,\mm+3\,\km=3\, (2\,\mm+\lm)>0.\nonumber
\end{align} 
Therefore, the set of inequalities \eqref{Prop} is implied from the smaller set:
\begin{empheq}[box=\widefbox]{align}
\me>0,\qquad\mh>0,\qquad \mc\geq0,\qquad  \ke+\kh>0,\qquad 2\,\mm+\lm>0.\label{cond1}
\end{empheq}
We note here that $3\,(2\,\me+\lle)\geq4\,\mm+3\,\ke\geq3\,(2\,\mm+\lm)$ because:
\begin{align}
3\,(2\,\me+\lle)&=4\,\me+3\,\ke \geq 4\,\mm+3\,\ke\geq 4\,\mm+3\,\km=3\,(2\,\mm+\lm).
\end{align}

\subsection{The $12\times12$ acoustic tensor for arbitrary direction}
We suppose that the space dependence of all introduced kinematic fields are limited to a direction defined by a unit vector $\xi$ which is the direction of propagation of the wave. Therefore, we look for solutions of:
\begin{align}
\rho\,  u_{,tt}=&\,\Div\left[2\,\me\, \sym\left(\nablau-\p\right)+2\,\mc \,\skew\left(\nablau-\p\right)+\lle\tr\left(\nablau-\p\right)\mathds{1}\right] , \nonumber \\\label{DynApp}
\eta  \p_{,tt}=&\,-\mLc  \Curl  \Curl \p+2\, \me \,\sym\left(\nablau-\p\right) +2 \,\mc\, \skew\left(\nablau-\p\right)\\&\hspace{2.625cm}+\lle\tr\left(\nablau-\p\right)\mathds{1}\ -\left[2\mh \sym \p +\lh \tr(\p)\mathds{1}\right],\nonumber 
\end{align}
in the form:
\begin{align}
u(x,t)&=\widehat{u} \, \underbrace{e^{i \left(k\langle \xi,\, x\rangle_{\R^3}-\,\omega \,t\right)}}_{s(x,t)\in\R/\mathbb{C}\ \text{scalar}}\,, &\widehat{u}&\in\mathbb{C}^{3}\,, &\lVert\xi\rVert^2&=1\,,\\\nonumber
\p(x,t)&=\widehat{\p} \,\underbrace{e^{i \left(k\langle \xi,\, x\rangle_{\R^3}-\,\omega \,t\right)}}_{s(x,t)\in\R/\mathbb{C}\ \text{scalar}}\,, &\widehat{\p}& \in\mathbb{C}^{3\times 3}\,.
\end{align} 
We start by remarking that considering $A,B\in\R^{3\times3}$ we have that:
\begin{align}
\Curl(A\cdot B)=L_B (\nabla A)+A\cdot \Curl(B),
\end{align}
therefore we obtain:
\begin{align}
\Curl_x(\widehat{\p}\cdot s(x,t))=\Curl(\widehat{\p}\cdot s(x,t)\cdot \id)=\widehat{\p}\cdot \Curl(s(x,t) \id),
\end{align}
where:
\begin{align}
\Curl(s(x,t) \id)&=\left(\begin{array}{ccc}
0 & \partial_3 s(x,t)& \partial_2 s(x,t) \\
-\partial_3 s(x,t) &0 & \partial_1 s(x,t)\\
\partial_2 s(x,t) &-\partial_1 s(x,t) &0
\end{array}\right)\in\so(3).
\end{align}
The derivatives of $s(x,t)$ can be evaluated considering:
\begin{align}
\nabla_x s(x,t) &=\left(\begin{array}{c} \partial_1 s(x,t)\\ \partial_2 s(x,t) \\
\partial_3 s(x,t)
\end{array}\right)=e^{i \left(k\langle \xi,\, x\rangle_{\R^3}-\,\omega \,t\right)}\left(\begin{array}{c} i\,k\,\xi_1\\ i\,k\,\xi_2 \\
i\,k\,\xi_3
\end{array}\right)=e^{i \left(k\langle \xi,\, x\rangle_{\R^3}-\,\omega \,t\right)}\,i\,k\,\xi=i\,k\,\xi\,s(x,t).
\end{align}
It can be noticed that:
\begin{align}
\Curl(s(x,t) \id)&=\mathrm{anti}(\nabla s(x,t))=e^{i \left(k\langle \xi,\, x\rangle_{\R^3}-\,\omega \,t\right)}\,i\,k\,\mathrm{anti}(\xi)=s(x,t)\,i\,k\,\mathrm{anti}(\xi).
\end{align}
Therefore, it is possible to evaluate the $\Curl\Curl\p$ as:
\begin{align}
\Curl\Curl(\widehat{\p}\, s(x,t))&=\Curl(\widehat{\p}\,\cdot\,\underbrace{\mathrm{anti}(\xi)}_{\in\so(3)}\,i\,k\, s(x,t))=i\,k\,\Curl([\widehat{\p}\,\cdot\,\mathrm{anti}(\xi)]\,\cdot \id s(x,t))=i\,k\,\widehat{\p}\,\cdot\,\mathrm{anti}(\xi)\,\Curl(\id s(x,t))\\
&=i\,k\,i\,k\,\widehat{\p}\,\cdot\,\mathrm{anti}(\xi)\,\cdot\,\mathrm{anti}(\xi)\,s(x,t)=-k^2\,\widehat{\p}\,\cdot\,\mathrm{anti}(\xi)\cdot\mathrm{anti}(\xi)\,e^{i \left(k\langle \xi,\, x\rangle_{\R^3}-\,\omega \,t\right)}.\nonumber
\end{align}
On the other hand, the second derivative of $\p$ with respect to time is:
\begin{align}
\p_{,tt}=\partial_t^2(\widehat{\p}e^{i \left(k\langle \xi,\, x\rangle_{\R^3}-\,\omega \,t\right)})=-\omega^2 \widehat{\p}e^{i \left(k\langle \xi,\, x\rangle_{\R^3}-\,\omega \,t\right)})=-\omega^2\widehat{\p}\,s(x,t).
\end{align}
Analogously for $u$ it is possible to evaluate the gradient and the derivatives with respect to time as:
\begin{align}
\nabla_{x} u&=i\,k\, s(x,t) \widehat{u}\otimes\xi,
&u_{,tt}&=-\omega^{2}\, \widehat{u}\, s(x,t).
\end{align}
The sym, skew and tr of $\nablau-\p$ can then be expressed as:
\begin{align}
\sym(\nablau-\p)&=\sym(i\,k\, \widehat{u}\otimes\xi -\widehat{\p})\, s(x,t)=(i\,k\,\sym( \widehat{u}\otimes\xi) -\sym\widehat{\p})\, s(x,t),\nonumber\\
\skew(\nablau-\p)&=\skew(i\,k\, \widehat{u}\otimes\xi -\widehat{\p})\, s(x,t)=(i\,k\,\skew( \widehat{u}\otimes\xi) -\skew\widehat{\p})\, s(x,t),\\
\tr(\nablau-\p)&=\tr(i\,k\, \widehat{u}\otimes\xi -\widehat{\p})\, s(x,t)=(i\,k\,\langle \widehat{u},\xi\rangle -\tr\widehat{\p})\, s(x,t).\nonumber
\end{align}
Therefore, we have:
\begin{align}
\Div\sym(\nablau-\p)&=\Div\left[(i\,k\,\sym( \widehat{u}\otimes\xi) -\sym\widehat{\p})\, \,s(x,t)\right]=(i\,k\,\sym( \widehat{u}\otimes\xi) -\sym\widehat{\p})\cdot \nabla_x\,s(x,t)\nonumber\\
&=(i\,k\,\sym( \widehat{u}\otimes\xi) -\sym\widehat{\p})\cdot (i\,k\,\xi\,s(x,t))=-(k^2\,\sym( \widehat{u}\otimes\xi)\cdot\,\xi +i\,k\,\sym\widehat{\p}\cdot\,\xi)\,s(x,t),\nonumber
\\
\Div\skew(\nablau-\p)&=\Div\left[(i\,k\,\skew( \widehat{u}\otimes\xi) -\skew\widehat{\p})\, s(x,t)\right]=(i\,k\,\skew( \widehat{u}\otimes\xi) -\skew\widehat{\p})\cdot \nabla_x\,s(x,t)\\
&=(i\,k\,\skew( \widehat{u}\otimes\xi) -\skew\widehat{\p})\cdot (i\,k\,\xi\,s(x,t))=-(k^2\,\skew( \widehat{u}\otimes\xi)\cdot\,\xi +i\,k\,\skew\widehat{\p}\cdot\,\xi)\,s(x,t),\nonumber\\
\Div(\tr(\nablau-\p)\,\id)&=\Div\left[\left((i\,k\,\langle \widehat{u},\xi\rangle -\tr\widehat{\p})\, \id\right) s(x,t)\right]=(i\,k\,\langle \widehat{u},\xi\rangle -\tr\widehat{\p})\id \cdot \nabla_x\,s(x,t)\nonumber\\
&=(i\,k\,\langle \widehat{u},\xi\rangle -\tr\widehat{\p})\id\cdot (i\,k\,\xi\,s(x,t))=-(k^2\,\langle \widehat{u},\xi\rangle +i\,k\,\tr\widehat{\p})\xi\,s(x,t)\nonumber.
\end{align}
Here, we have used the relationship:
\begin{align}
\Div[B\  s(x,t)]=\underbrace{\Div[B ]}_{=0}s(x,t)+B\cdot \nabla_x s(x,t),
\end{align}
where $B\in\R^{3\times 3}$ and $s(x,t)$ is a scalar. With all the formulas obtained it is possible to write \eqref{DynApp} simplifying $s(x,t)$ everywhere as:
\begin{align}
-\rho\, \omega^{2}\, \widehat{u}=&-\big[2\,\me\, (k^2\,\sym( \widehat{u}\otimes\xi)\cdot\,\xi +i\,k\, \sym\widehat{\p}\cdot\,\xi))+2\,\mc\,(k^2\,\skew( \widehat{u}\otimes\xi)\cdot\,\xi +i\,k\,\skew\widehat{\p}\cdot\,\xi)\nonumber \\
&+\lle(k^2\,\langle \widehat{u},\xi\rangle +i\,k\,\tr\widehat{\p})\,\xi\big] , \nonumber \\
-\eta\, \omega^2\widehat{\p}=&\,\mLc k^2\,\widehat{\p}\,\mathrm{anti}(\xi)\cdot\mathrm{anti}(\xi)+2\, \me \,(i\,k\,\sym( \widehat{u}\otimes\xi) -\sym\widehat{\p}) +2 \,\mc\, (i\,k\,\skew( \widehat{u}\otimes\xi) -\skew\widehat{\p})\\&+\lle(i\,k\,\langle \widehat{u},\xi\rangle -\tr\widehat{\p})\mathds{1} -\left[2\mh \sym \widehat{\p} +\lh \tr(\widehat{\p})\mathds{1}\right],\nonumber 
\end{align}
or analogously:
\begin{align}
- \rho\, \omega^{2}\,\widehat{u}+k^2 (2\,\me\,\sym( \widehat{u}\otimes\xi)\cdot\,\xi+2\,\mc\,\,\skew( \widehat{u}\otimes\xi)\cdot\,\xi+\lle\,\langle \widehat{u},\xi\rangle\,\xi)\nonumber\\+i\,k\,(2\,\me \sym\widehat{\p}\cdot\,\xi+2\,\mc \skew\widehat{\p}\cdot\,\xi+\lle\,\tr\widehat{\p}\,\xi)&=\,0 , \nonumber \\
-\eta\, \omega^2\widehat{\p}-\mLc k^2\,\widehat{\p}\,\mathrm{anti}(\xi)\cdot\mathrm{anti}(\xi)+2(\me+\mh) \sym \widehat{\p}+2\mc\,\skew \widehat{\p} +(\lle+\lh) \tr(\widehat{\p})\mathds{1}\\-\,2\, \me \,i\,k\,\sym( \widehat{u}\otimes\xi) -2 \,\mc\, i\,k\,\skew( \widehat{u}\otimes\xi)-\lle i\,k\,\langle \widehat{u},\xi\rangle \mathds{1}&=0.\nonumber 
\end{align}
At given $\xi\in\R^3$, this is a linear system in $(\widehat{u},\widehat{\p})\in\mathbb{C}^{12}$ which can be written in $12\times12$ matrix format as:
\begin{align}
\Vast(\quad \widetilde{A}(\xi,\omega,k)\quad \Vast) \left(\begin{array}{c}\widehat{u}_1\\\widehat{u}_2\\\widehat{u}_3\\\widehat{\p}_{11}\\\widehat{\p}_{12}\\\widehat{\p}_{13}\\\widehat{\p}_{21}\\\widehat{\p}_{22}\\\widehat{\p}_{23}\\\widehat{\p}_{31}\\\widehat{\p}_{32}\\\widehat{\p}_{33}\end{array}\right)&=0,&
\Vast(\quad \widetilde{\mathcal{B}}(\xi,k)-\omega^2 \id\quad \Vast) \left(\begin{array}{c}\widehat{u}_1\\\widehat{u}_2\\\widehat{u}_3\\\widehat{\p}_{11}\\\widehat{\p}_{12}\\\widehat{\p}_{13}\\\widehat{\p}_{21}\\\widehat{\p}_{22}\\\widehat{\p}_{23}\\\widehat{\p}_{31}\\\widehat{\p}_{32}\\\widehat{\p}_{33}\end{array}\right)&=\left(\begin{array}{c}0\\0\\0\\0\\0\\0\\0\\0\\0\\0\\0\\0\end{array}\right).\label{sys12}
\end{align}
Here, $\widetilde{\mathcal{B}}(\xi,k)$ is the $12\times12$ acoustic tensor. The columns of $\widetilde{\mathcal{A}}$ are:
\begin{align}\nonumber
\widetilde{A}_{i1}&=\left(\begin{array}{ccc}
\rho \, \omega^2-k^2 (\lle+2\me)\xi_1^2-k^2(\mc+\me)(\xi_2^2+ \xi_3^2)\\
-k^2 (\lle-\mc+\me)\xi_1 \xi_2 \\
-k^2 (\lle-\mc+\me) \xi_1 	\xi_3 \\
i k (\lle+2\me)\xi_1\\
i k (\mc+\me) \xi_2 \\
i k (\mc+\me) \xi_3 \\
- i k (\mc-\me) \xi_2\\
i k \lle \xi_1 \\
0 \\
-i k (\mc-\me)\xi_3\\
0\\
i k \lle \xi_1
\end{array}\right), \
&\widetilde{A}_{i2}&=\left(
\begin{array}{c}
-k^2 (\lle-\mc+\me)\xi_1 \xi_2\\
\rho \, \omega^2-k^2 (\lle+2\me)\xi_2^2-k^2(\mc+\me)(\xi_1^2+ \xi_3^2)\\-k^2 (\lle-\mc+\me) \xi_2 \xi_3\\
i k \lle \xi_2 \\
- i k (\mc-\me) \xi_1\\
0\\
i k (\mc+\me)\xi_1\\
i k (\lle+2\me) \xi_2\\
i k (\mc +\me) \xi_3\\
0\\
-i k (\mc-\me) \xi_3 \\
i k \lle \xi_2
\end{array}\right),
\end{align} \begin{align}
\nonumber
\widetilde{A}_{i3}&=\left(
\begin{array}{c}
-k^2 (\lle-\mc+\me) \xi_1 \xi_3\\
-k^2 (\lle-\mc+\me) \xi_2 \xi_3\\
\rho \, \omega^2-k^2 (\lle+2\me)\xi_3^2-k^2(\mc+\me)(\xi_1^2+ \xi_2^2) \\
i k \lle \xi_3\\
0\\
- i k (\mc - \me) \xi_1 \\
0\\
i k \lle \xi_3\\
- i k (\mc-\me)\xi_2\\
i k (\mc+\me) \xi_1 \\
i k (\mc+\me )\xi_2\\
i k (\lle+2\me) \xi_3
\end{array}\right),\ &\widetilde{A}_{i4}&=\left(
\begin{array}{c}
- i k (\lle+2\me)\xi_1\\
-i k \lle \xi_2\\
- i k \lle \xi_3\\
\eta \, \omega^2-(2(\me+\mh)+\lle+\lh)-k^2\mLc (\xi_2^2+\xi_3^3)\\
k^2 \mLc \xi_1 \xi_2\\
k^2 \mLc \xi_1 \xi_3\\
0\\
-(\lle+\lh)\\
0\\
0\\
0\\
-(\lle+\lh)
\end{array}\right),
\end{align}\begin{align}
\nonumber
\widetilde{A}_{i5}&=\left(
\begin{array}{c}
-i k (\mc+ \me) \xi_2\\
i k (\mc-\me) \xi_1\\
0 \\
k^2 \mLc \xi_1\xi_2\\
\eta \, \omega^2-(\mc+\me+\mh)-k^2\mLc (\xi_1^2+\xi_3^2)\\
k^2 \mLc \xi_1 \xi_2\\
0\\
0\\
0\\
0\\
0
\end{array}\right),\ 
&\widetilde{A}_{i6}&=\left(
\begin{array}{c}
-i k (\mc+\me)\xi_3\\
0\\
i k (\mc-\me) \xi_1\\
k^2\mLc\xi_1\xi_3 \\
k^2 \mLc \xi_2 \xi_3\\
\eta \, \omega^2-(\mc+\me+\mh)-k^2\mLc (\xi_1^2+\xi_2^2)\\
0\\
0\\
0\\
\mc-\me-\mh\\
0\\
0
\end{array}\right),
\end{align}\begin{align}
\nonumber
\widetilde{A}_{i7}&=\left(
\begin{array}{c}
i k (\mc-\me)\xi_2\\
-i k (\mc+\me)\xi_1\\
0\\
0\\
\mc-\me-\mh\\
0\\
\eta \, \omega^2-(\mc+\me+\mh)-k^2\mLc (\xi_2^2+\xi_3^2)\\
k^2 \mLc \xi_1\xi_2\\
k^2 \mLc \xi_1	\xi_3\\
0\\
0\\
0
\end{array}\right),\
&\widetilde{A}_{i8}&=\left(
\begin{array}{c}
-i k \lle \xi_1\\
-i k (2\me+\lle)\xi_2\\ 
-i k \lle \xi_3\\
- \lle-\lh\\
0\\
0\\
k^2 \mLc \xi_1 \xi_2\\
\eta \, \omega^2-(\mc+\me+\mh)-k^2\mLc (\xi_1^2+\xi_3^2)\\
k^2 \mLc \xi_2 \xi_3\\
0\\
0\\
-\lle-\lh
\end{array}\right),
\end{align}\begin{align}
\nonumber
\widetilde{A}_{i9}&=\left(
\begin{array}{c}
0\\
-i k (\mc+\me) \xi_3\\
i k (\mc+\me)\xi_2\\
0\\
0\\
0\\
k^2 \mLc \xi_1 \xi_3\\
k^2 \mLc \xi_2 \xi_3\\
\eta \, \omega^2-(\mc+\me+\mh)-k^2\mLc (\xi_1^2+\xi_2^2)\\
0\\
\mc-\me-\mh\\
0
\end{array}\right),\ 
&\widetilde{A}_{i10}&=\left(\begin{array}{c}
i k (\mc-\me) \xi_3\\
0\\
- i k (\mc+\me)\xi_1\\
0\\
0\\
\mc-\me-\mh\\
0\\
0\\
0\\
\eta \, \omega^2-(\mc+\me+\mh)-k^2\mLc (\xi_2^2+\xi_3^2)\\
k^2 \mLc \xi_1\xi_2\\
k^2 \mLc \xi_1\xi_3
\end{array}\right),
\end{align}\begin{align}
\nonumber
\widetilde{A}_{i11}&=\left(
\begin{array}{c}
0\\
i k (\mc-\me) \xi_3\\
- i k (\mc+\me)\xi_2\\
0\\
0\\
0\\
0\\
0\\
\mc-\me-\mh\\
k^2 \mLc \xi_1\xi_2\\
\eta \, \omega^2-(\mc+\me+\mh)-k^2\mLc (\xi_1^2+\xi_3^2)\\
k^2 \mLc \xi_2\xi_3
\end{array}\right),\ 
&\widetilde{A}_{i12}&=\left(
\begin{array}{c}
-i k \lle \xi_1\\
-i k \lle \xi_2\\
- i k (\lle+2\,\me)\xi_3\\
-\lle-\lh\\
0\\
0\\
0\\
-\lle-\lh\\
0\\
k^2 \mLc \xi_1\xi_3\\
k^2 \mLc \xi_2\xi_3\\
\eta \, \omega^2-(\mc+\me+\mh)-k^2\mLc (\xi_1^2+\xi_2^2)
\end{array}\right).
\end{align}
It is clear that even with the aid of up-to-date computer algebra systems, it is practically impossible to determine positive-definiteness of the $12\times12$ acoustic tensor $\widetilde{\mathcal{B}}$ in dependence of the given material parameters. In the main body of our paper we succeed by choosing immediately the propagation direction $\xi=e_1$ and by considering a set of new variables \eqref{Decom}. This allows us to obtain a certain pre-factorization of $\widetilde{\mathcal{B}}(e_1,k)$ in $3\times3$ blocks. Since the formulation is isotropic, choosing $\xi=e_1$ is no restriction, as argued before.

\end{document}